\documentclass[11pt]{article}

\usepackage[latin1]{inputenc} 
\usepackage[nosort]{cite}
\usepackage[bulletsep]{collref}
\usepackage{graphicx}
\usepackage{color} 
\usepackage{bbm}
\usepackage{amsmath}
\usepackage{amssymb}
\usepackage{subfigure}
\usepackage{array}
\usepackage[bookmarks=true,hyperfigures=true]{hyperref}

\usepackage[a4paper,includeheadfoot,hmargin=2.5cm,vmargin=2cm,headheight=14pt,headsep=2.5ex,footskip=5.5ex]{geometry}

\providecommand{\hypersetup}[1]{}

\providecommand{\pdfbookmark}[3][]{}

\hypersetup{plainpages=false}
\hypersetup{pdfpagemode=UseOutlines}
\hypersetup{bookmarksnumbered=true}
\hypersetup{bookmarksopen=true}
\hypersetup{pdfstartview=FitH}
\hypersetup{colorlinks=false}
\hypersetup{citebordercolor={.6 .9 .6}}
\hypersetup{urlbordercolor={.7 .8 1}}
\hypersetup{linkbordercolor={1 .7 .7}}
\hypersetup{pdfborder={0 0 .5}}

\numberwithin{equation}{section}

\let\oldbfseries=\bfseries
\let\oldmdseries=\mdseries
\let\oldnormalfont=\normalfont
\renewcommand{\bfseries}{\oldbfseries\boldmath}
\renewcommand{\mdseries}{\oldmdseries\unboldmath}
\renewcommand{\normalfont}{\oldnormalfont\unboldmath}

\makeatletter
\let\old@makecaption=\@makecaption
\def\@makecaption{\small\old@makecaption}
\makeatother

\newcommand{\superN}{\mathcal{N}}
\newcommand{\Smat}{\mathcal{S}}
\newcommand{\Tmat}{\mathcal{T}}
\newcommand{\Action}{\mathcal{S}}
\newcommand{\Lagr}{\mathcal{L}}

\newcommand{\unit}{\mathbbm{1}}

\newcommand{\order}{\mathcal{O}}

\newcommand{\trans}{{\scriptscriptstyle\mathrm{T}}}

\newcommand{\Reals}{\mathbbm{R}}

\newcommand{\Sphere}{S}  
\newcommand{\AdS}{\mathrm{AdS}}

\DeclareMathOperator*{\Res}{Res}

\ifx\genfrac\sdflkaj

\else

\fi
\newcommand{\sfrac}[2]{{\textstyle\frac{#1}{#2}}}
\newcommand{\half}{\sfrac{1}{2}}

\newcommand{\Half}{\frac{1}{2}}
\newcommand{\iHalf}{\frac{i}{2}}
\newcommand{\Quarter}{\frac{1}{4}}

\newcommand{\alg}[1]{\mathfrak{#1}}
\newcommand{\grp}[1]{\mathrm{#1}}

\newcommand{\grSU}{\grp{SU}}
\newcommand{\grSO}{\grp{SO}}

\newcommand{\lrbrk}[1]{\left(#1\right)}
\newcommand{\bigbrk}[1]{\bigl(#1\bigr)}
\newcommand{\Bigbrk}[1]{\Bigl(#1\Bigr)}

\newcommand{\lrsbrk}[1]{\left[#1\right]}
\newcommand{\bigsbrk}[1]{\bigl[#1\bigr]}
\newcommand{\Bigsbrk}[1]{\Bigl[#1\Bigr]}
\newcommand{\biggsbrk}[1]{\biggl[#1\biggr]}
\newcommand{\Biggsbrk}[1]{\Biggl[#1\Biggr]}
\newcommand{\ket}[1]{\mathopen{|}#1\mathclose{\rangle}}
\newcommand{\bra}[1]{\mathopen{\langle}#1\mathclose{|}}

\newcommand{\lrabs}[1]{\left|#1\right|}
\newcommand{\abs}[1]{{|#1|}}

\newcommand{\eval}[1]{#1|}

\newcommand{\levi}{\epsilon}
\newcommand{\comma}{\quad,\quad}

\newcommand{\nn}{\nonumber}
\newcommand{\nln}{\nonumber\\}
\newcommand{\nl}[1][0pt]{\nonumber\\[#1]&\hspace{-4\arraycolsep}&\mathord{}}

\newcommand{\earel}[1]{\mathrel{}&\hspace{-2\arraycolsep}#1\hspace{-2\arraycolsep}&\mathrel{}}
\newcommand{\eq}{\earel{=}}

\def\[{\begin{equation}}
\def\]{\end{equation}}
\def\<{\begin{eqnarray}}
\def\>{\end{eqnarray}}

\makeatletter
\def\mr@ignsp#1 {\ifx\:#1\@empty\else #1\expandafter\mr@ignsp\fi}%
\newcommand{\multiref}[1]{\begingroup
\xdef\mr@no@sparg{\expandafter\mr@ignsp#1 \: }%
\def\mr@comma{}%
\@for\mr@refs:=\mr@no@sparg\do{\mr@comma\def\mr@comma{,}\ref{\mr@refs}}%
\endgroup}
\makeatother

\newcommand{\hypref}[2]{\ifx\href\asklfhas #2\else\href{#1}{#2}\fi}

\newcommand{\secref}[1]{Sec.~\multiref{#1}}

\newcommand{\appref}[1]{App.~\multiref{#1}}

\newcommand{\figref}[1]{Fig.~\multiref{#1}}
\renewcommand{\eqref}[1]{(\multiref{#1})}

\ifx\href\asklfhas\newcommand{\href}[2]{#2}\fi

\newcommand{\Bsi}{\Upsilon}
\newcommand{\lAA}{{a}}

\newcommand{\laa}{{\alpha}}

\newcommand{\lBB}{{b}}

\newcommand{\lbb}{{\beta}}

\newcommand{\lCC}{{c}}

\newcommand{\lcc}{{\gamma}}

\newcommand{\lDD}{{d}}

\newcommand{\ldd}{{\delta}}

\newcommand{\eps}{\varepsilon}

\newcommand{\Smatrix}{S}  
\newcommand{\smatrix}{\mathsf{S}}          

\let\vecarrow=\vec
\renewcommand{\vec}[1]{\mathbf{#1}}
\newcommand{\vecsigma}{\boldsymbol{\sigma}}

\begin{document}

\thispagestyle{empty}
\begin{flushright}\footnotesize
\texttt{UUITP-23/12}\\
\texttt{AEI-2012-082}%
\end{flushright}
\vspace{1cm}

\begin{center}%
{\Large\textbf{\mathversion{bold}%
Worldsheet Form Factors in AdS/CFT
}\par}

\vspace{1.5cm}

\textrm{Thomas Klose$^{a}$ and Tristan McLoughlin$^{b}$} \vspace{8mm} \\
\textit{%
$^a$ Department of Physics and Astronomy, Uppsala University \\
SE-75108 Uppsala, Sweden \\
$^b$ Max-Planck-Institut f\"ur Gravitationsphysik, Albert-Einstein-Institut, \\
Am M\"uhlenberg 1, D-14476 Potsdam, Germany
} \\
\texttt{\\ thomas.klose@physics.uu.se, tmclough@aei.mpg.de}

\par\vspace{14mm}

\textbf{Abstract} \vspace{5mm}

\begin{minipage}{14cm}
We formulate a set of consistency conditions appropriate to worldsheet form factors in the massive, integrable but non-relativistic, light-cone gauge fixed AdS$_5 \times$S$^5$ string theory. We then perturbatively verify that these conditions hold, at tree level in the near-plane-wave limit and to one loop in the near-flat (Maldacena-Swanson) limit, for a number of specific cases. We further study the form factors in the weakly coupled dual description, verifying that the relevant conditions naturally hold for the one-loop Heisenberg spin-chain. Finally, we note that
the near-plane-wave expressions for the form factors, when further expanded in small momentum or, equivalently, large charge density, reproduce the thermodynamic limit of the spin-chain results at leading order.
\end{minipage}

\end{center}

\newpage
\tableofcontents

\vspace{10mm}
\hrule
\vspace{5mm}

\section{Introduction}

The on-shell properties of massive, integrable, two-dimensional quantum field theories can be completely characterized by their two-particle S-matrix. The exact expression for this S-matrix can in many cases be calculated by making use of unitarity, crossing symmetry and the underlying global symmetries of the theory, see e.g. \cite{Zamolodchikov:1978xm, Dorey:1996gd} for reviews. Even more interestingly, knowledge of the S-matrix can also be used to calculate off-shell quantities. For instance, form factors, i.e.\ the matrix elements of the local operators, $\mathcal{O}$, in the basis of asymptotic in- and out-particle states,
\<
  \bra{\mathrm{out}} \mathcal{O}(\vec{x}) \ket{\mathrm{in}} \; ,
\>
are largely determined by unitarity, analyticity and---for Lorentz invariant theories---relativistic symmetry. These properties allow for the formulation of a set of consistency conditions, the so-called form factor axioms, which involve the S-matrix. These axioms were first written down by \cite{Weisz:1977ii, Karowski:1978vz} and further developed in many works, e.g. \cite{Kirillov:1987jp} (see \cite{Smirnov:1992vz} for a thorough exposition). From these form factors, one can then further build correlation functions
\<
  \bra{\Omega} \mathcal{O}_1(\vec{x}) \mathcal{O}_2(\vec{y}) \cdots \ket{\Omega}
\>
of the corresponding operators. The correlation functions are expressible as sums of products of form factors by inserting complete sets of scattering states between the operators.

We will be interested in the calculation of form factors for the AdS$_5$ $\times$ S$^5$ string worldsheet theory and we will concentrate on the light-cone gauge fixed version which is a massive integrable theory albeit not Lorentz invariant. A conjectured exact S-matrix, for which there is significant evidence, exists \cite{Arutyunov:2004vx, Beisert:2005tm,Beisert:2006ez,Beisert:2006ib} (see the reviews \cite{Arutyunov:2009ga} and \cite{Beisert:2010jr} for an overview and appropriate references). However, the absence of Lorentz invariance and a generally more complicated analytic structure of the theory, means that the relativistic axioms need to be slightly generalized before the usual form factor program can be developed for the AdS$_5$ $\times$ S$^5$ worldsheet theory. Nonetheless, in spite of this difference the calculation of form factors in sine-Gordon/massive Thirring and sinh-Gordon models provides a very useful guideline. We will thus briefly review the standard form factor bootstrap following in particular \cite{Fring:1992pt, Babujian:1998uw, Babujian:2001xn}, which make clear the connection to the Lagrangian description of the theory, before reviewing the exact S-matrix of the worldsheet integrable model. We will then propose such a set of consistency conditions for the worldsheet form factors which can hopefully be used as a set of axioms for off-shell local operators in the AdS$_5$ $\times$ S$^5$ theory.

We will check these conditions perturbatively, in various limits, for a number of different configurations. At large values of the effective string tension, $\sqrt{\lambda}\gg1$, the worldsheet theory is weakly coupled and form factors can be calculated by standard Feynman diagrammatic methods and by using LSZ reduction to relate worldsheet time-ordered correlation functions to asymptotic states. Such perturbative calculations have previously been performed for the tree-level worldsheet S-matrix  \cite{Klose:2006zd} and here we will use the same method to calculate the tree-level one- and three-particle form factors where the local operator is one of the complex scalar fields of the theory. We then turn to the Maldacena-Swanson or near-flat limit \cite{Maldacena:2006rv}, which can be viewed as a truncation to the sector of excitations which are highly boosted along one of the worldsheet directions. This significantly simplifies the worldsheet theory and the S-matrix has been calculated to one- and two-loop \cite{Klose:2007wq, Klose:2007rz} order, which in addition to providing evidence for the conjectured exact S-matrix validates the consistency of this limit. Indeed, in this limit one can even prove factorization of the one-loop three-particle S-matrix \cite{Puletti:2007hq}, one of the few direct checks of quantum integrability for the worldsheet theory. We will use the near-flat limit to  calculate matrix elements of a single complex scalar but extend our calculations to one-loop, where the analytical structure is more non-trivial, and to the case where the operator is a composite of two scalars. At this order we check the analogue of Watson's equations, \cite{Watson:1954uc}, and study the behavior of the one-particle poles. 

The worldsheet theory is known to posses a rich spectrum of bound states \cite{Hofman:2006xt, Dorey:2006dq} which will give rise to poles in the form factors, however these states cannot be seen in perturbation theory about the trivial vacuum. To gain some insight, we consider the opposite limit
of small 't Hooft coupling, $\lambda \ll 1$, where the integrable model corresponds to an integrable spin-chain. In the so-called $\alg{su}(2)$ sector corresponding to the single complex worldsheet scalar, this is just the Heisenberg XXX model. The form factors correspond to the matrix elements of spin-chain operators; the study of such objects, and spin-chain correlation functions in general, is another well developed, though still challenging, area in integrable models (see e.g. \cite{1993qism.book.korepin}). However, one only needs the relatively pedestrian coordinate Bethe ansatz description of spin-chain states to immediately see that the  form factor axioms, including the bound state axiom, 
hold in this limit for the cases we consider.

For large charge density, or infinite length, a direct comparison can be made between the spectrum of the string theory and that of the 
Landau-Lifshitz theory describing the long wavelength behavior of the spin-chain \cite{Kruczenski:2003gt, Kruczenski:2004kw}. Here we show 
that by choosing the appropriate light-cone gauge of the near-plane-wave expansion and after making an appropriate field redefinition we can also find a match for the form factors to leading order in a further small momentum expansion. This match is quite analogous to that found
between string  and one-loop spin-chain energies \cite{Parnachev:2002kk,Callan:2003xr,Callan:2004uv} which suggests that it will fail at sufficiently high order but also
that the non-trivial interpolation between weak and strong coupling may be extended off-shell. In summary, our calculations demonstrate the feasibility of applying the 
form factor program to the string worldsheet theory which should provide a new direction toward a complete solution of the model.

\section{Form Factors}

\subsection{Definition of Form Factors}

We start with a generic two-dimensional theory where each external particle is characterized by a two-momentum, $\vec{p}=(\epsilon,p)$, and an internal particle flavor index $i$. In the worldsheet theory such an index will run over the transverse bosonic and fermionic worldsheet fields of the light-cone gauge fixed theory. Consider any local operator $\mathcal{O}(\vec{x})$, where the operator is some composite of the fundamental fields and their derivatives located at some point $\vec{x}=(\tau, \sigma)$. We can define the generalized form factor in terms of the matrix elements between scattering states
\< \label{eqn:ff-def-indices}
{}^{i'_m,\dots, i'_1}\bra{ p'_m,\dots, p'_1} \mathcal{O}(\vec{x}) \ket{ p_1,\dots, p_n }_{i_1,\dots, i_n}
  \eq e^{i(\vec{p}'_{1} + \ldots + \vec{p}'_{m} - \vec{p}_1 - \ldots - \vec{p}_n )\cdot \vec{x}} \nl
      \times F^{\mathcal{O}; i'_m,\dots, i'_1}_{i_1,\dots, i_n}(p'_m,\dots, p'_1|p_1,\dots,p_n) \; .
\>
Here, the attribute ``generalized'' refers to two facts, namely that there are particles (or anti-particles) in both external state and that the momenta are not ordered. We will now see that it is sufficient to consider more ``specialized'' form factors.

For a relativistic theory we can use crossing to relate a generic form factors to matrix elements between the vacuum, $\ket{\Omega}$, and a single external $n$-particle state. While the string worldsheet theory in light-cone gauge is not Lorentz invariant, there nonetheless exists a notion of crossing and so we can similarly focus on the same matrix elements
\< \label{eqn:ff-def-vac-coord}
\bra{\Omega} \mathcal{O}(\vec{x}) \ket{ p_1,\dots,p_n }_{ i_1,\dots, i_n}
  \eq e^{ - i (\vec{p}_1 + \ldots + \vec{p}_n ) \cdot \vec{x}} F^{\mathcal{O}}_{\underline{i}}(p_{\underline i}) \; ,
\>
or for operators in momentum space $\tilde{ \mathcal{O}}(\vec{q})=\int d^2x~ e^{i \vec{q} \cdot \vec{x} } \mathcal{O}(\vec{x})$:
\< \label{eqn:ff-def-vac-mom}
 \bra{\Omega} \tilde{ \mathcal{O}}(\vec{q}) \ket{ p_1,\dots, p_n }_{ i_1,\dots, i_n}
  \eq (2\pi)^2 \delta^{(2)}\bigbrk{ \vec{q}  - \vec{p}_1 - \ldots - \vec{p}_n } F^{\mathcal{O}}_{\underline{i}}(p_{\underline i}) \; .
\>
For convenience we have adopted the notations $\underline{i}=\{i_1, \dots, i_n\}$ and $p_{\underline i}=\{p_1,\dots,p_n\}$.

The external states in \eqref{eqn:ff-def-indices}--\eqref{eqn:ff-def-vac-mom} are the conventional scattering states with positive energy wave-functions. However, in the relevant literature, it is customary to associate with the terms ``in''- and ``out''-scattering states a specific ordering of the momenta. In-scattering states, $\ket{p_1,\dots,p_n}_{ i_1, \dots, i_n}^{\rm (in)}$, are defined as incoming states, $\ket{p_1,\dots,p_n}_{ i_1, \dots, i_n}$, where $p_1>p_2>\dots>p_n$ and out-scattering states, $\ket{p_1,\dots,p_n}_{ i_1, \dots, i_n}^{\rm (out)}$, as outgoing states, $\ket{p_n,\dots,p_1}_{ i_n, \dots, i_1}$, where also $p_1>p_2>\dots>p_n$. As it plays an obviously key role in the study of form factors, let us note that the usual two-dimensional scattering matrix can be defined in this notation as 
\<
\label{eq:smatrix_def}
\ket{p_1,\dots, p_n}^{\rm (in)}_{i_1,\dots, i_n}=\ket{p_1,\dots, p_n}^{\rm (out)}_{\tilde{i}_1,\dots, \tilde{i}_n}\Smat^{\tilde{i}_1,\dots,\tilde{i}_{n}}_{i_1,\dots,i_n}(p_1,\dots, p_n)~.
\>
This formula in fact defines the S-matrix for a specific configuration of the particle momenta, it is defined by analytical continuation for other configurations which we discuss in the next section. 

We can now finally introduce the auxiliary functions, or form factors, $f^{\mathcal{O}}_{\underline i}(p_{\underline i} )$, defined to be equal to the matrix elements of operators at the origin, ${\cal O}={\cal O}(0)$, for the ``in''-ordering of momenta,
\< \label{eqn:def-aux-ff}
  f^{\mathcal{O}}_{\underline i}(p_{\underline i} ) = \bra{\Omega} \mathcal{O} \ket{ p_1,\dots, p_n }^{\rm (in)}_{ i_1, \dots, i_n}~,
\>
and extended to all other orderings by analytical continuation.

\subsection{Review of Relativistic Case}

The analytical properties of the observables such as the S-matrix or form factors are key in properly defining them and, where it is possible, in determining their exact expressions. We will first briefly review the standard Lorentz invariant case, essentially repeating the discussion in \cite{Babujian:1998uw}, before discussing the $\AdS_5\times\Sphere^5$ string case. 

\begin{figure}
\begin{center}
\includegraphics[scale=0.6]{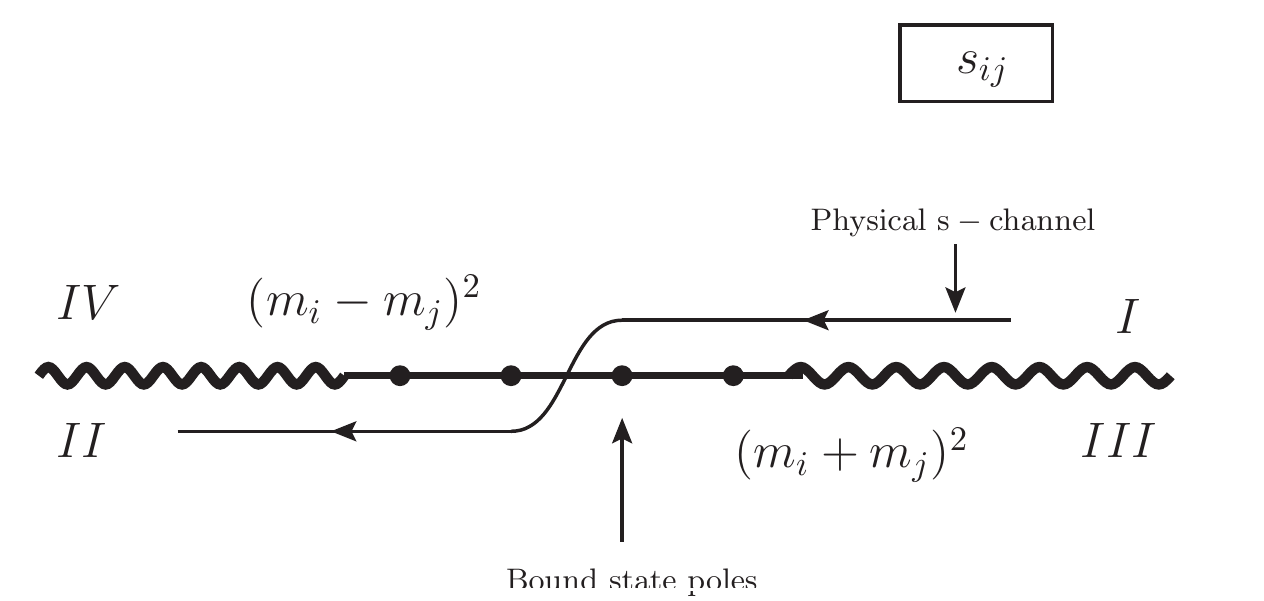}
\end{center}
\caption{Crossing for a Lorentz Invariant S-matrix.}
\label{fig:Crossing_LI_Smatrix}
\end{figure}

In a Lorentz invariant theory the S-matrix is naturally be thought of as a complex function of the Lorentz invariants, e.g. $s_{ij}=(\vec{p}_i+\vec{p}_j)^2$ with  $p_i$ and $p_j$ the momenta of any two incoming particles with mass $m_i$ and $m_j$, respectively. For an integrable theory because there is only elastic scattering the S-matrix will have only two branch cuts\footnote{Besides these branch cuts, the S-matrix generically has poles associated to the exchange of bound states or a number of on-shell particles. While the exchange of particles that are not bound leads to branch cuts in higher dimensions, in two dimensions these processes lead to double or higher-order poles \cite{Coleman:1978kk,Dorey:1996gd}.},  one in the s-channel, that is the kinematical region where $s_{ij}> (m_i+m_j)^2$, and one in the t-channel  where $s_{ij}< (m_i-m_j)^2$, which gives rise to four distinct regions, see \figref{fig:Crossing_LI_Smatrix}. The physical s-channel and t-channel regions are labeled by I and II, respectively. For example, for the two-to-two-particle scattering the physical region, i.e. positive energies and real momenta, corresponds to the boundary value of this analytic function
\<
S(1,2\rightarrow 3,4)=\lim_{ \epsilon\rightarrow 0+} S(s_{12}+i \varepsilon)~,~~ ~{\rm with}~ s_{12}>(m_1+m_2)^2~.
\>  
This corresponds to the usual $i\varepsilon$-prescription in perturbative calculations. 
The crossing transformation, which corresponds to the exchange of physical in- and out-waves, is shown by the arrowed line in \figref{fig:Crossing_LI_Smatrix},
 and is given by  $s_{ij}+i \varepsilon\leftrightarrow t_{ij}- i \varepsilon$, where $t_{ij}=(\vec{p}_i-\vec{p}_j)^2$.

\begin{figure}
\begin{center}
\includegraphics[scale=0.6]{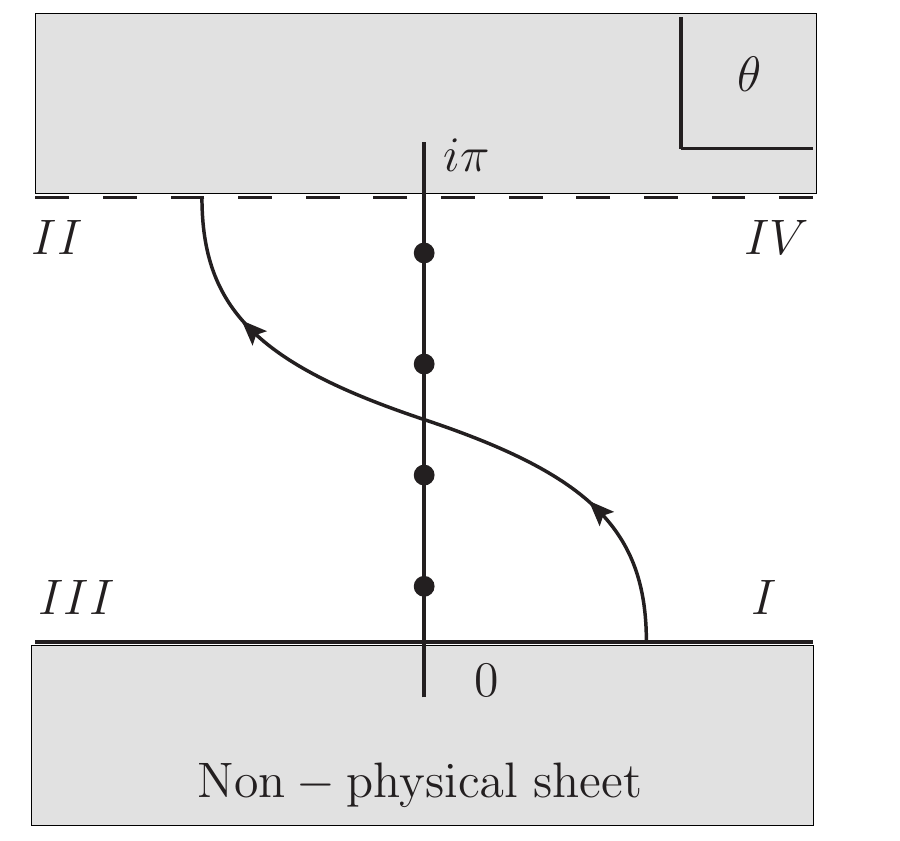}
\end{center}
\caption{Crossing for a Lorentz Invariant S-matrix in terms of rapidity variables.}
\label{fig:Crossing_LI_Smatrix_rapidity}
\end{figure}

As is standard we can introduce the uniformizing parameterization i.e. rapidities, $\epsilon_i=m_i \cosh \theta_i$,  $p_i=m_i \sinh \theta_i$, 
and consider the S-matrix as a function of the rapidity difference, $\theta=|\theta_i-\theta_j|$. For the integrable theory the S-matrix  is now a meromorphic function defined on the strip
$0\leq {\rm Im}~ \theta\leq \pi$. The s-channel cut is mapped to the Im $\theta = \pi$ line and the t-channel cut to the Im $\theta=0$ line. 
The crossing relation is now given by the transformation $\theta\leftrightarrow i\pi -\theta$, see \figref{fig:Crossing_LI_Smatrix_rapidity}, which 
acts on the two-particle S-matrix as 
\<
S_{i_1i_2}^{i'_1i'_2}(i\pi-\theta)={C}^{-1}_{i_1j_1}~S^{j_1i'_2}_{j'_1i_2}(\theta)~C^{j'_1i'_1}
\>
where $C^{i_1 j_1}$ is the charge conjugation matrix involved in the exchange of particles with antiparticles.

The generalized form factors with $n$ particles in a Lorentz invariant theory can also be thought of as analytic functions of the Lorentz invariants $s_{ij}$ and $t_{ij}$. The boundary value of this function corresponds to the matrix element with all physical incoming particles, 
\<
F^{\cal O}_{\underline i} (\{s_{ij}+i \varepsilon\}_{1\leq i< j\leq n})=\bra{\Omega} \mathcal{O} \ket{ p_1,\dots, p_n }_{ i_1, \dots, i_n}~, 
\>
and where again the appropriate limit corresponds to the $i\varepsilon$-prescription in perturbation theory. Matrix elements with outgoing particles
can again be reached by crossing transformations. 

As for the scattering amplitudes, the form factors possess branch cuts. 
Considering the simplest generalized form factor with two external particles both with mass $m$, it is straightforward to see 
that there is a branch cut for $s_{12}=(p_1+p_2)^2>4m^2$ such that 
\<
F^{\cal O}_{i_1i_2}(s_{12}+i \varepsilon)=F^{\cal O}_{i'_1i'_2}(s_{12}-i \varepsilon)S^{i'_1i'_2}_{i_1i_2}(p_1,p_2)~,
\>
while there is no cut in the $t_{12}=(p_1-p_2)^2$ channel 
\<
F^{\cal O}_{i_1i_2}(t_{12}-i \varepsilon)=F^{\cal O}_{i_1i_2}(t_{12}+i \varepsilon)~.
\>
That is, unlike for the S-matrix, there is no branch cut in the t-channel. The extension of these relations to general form factors with arbitrary numbers of external particles are known as Watson's equations \cite{Watson:1954uc} and play a central role in the theory of form factors in integrable models. 

\begin{figure}
\begin{center}
\includegraphics[scale=0.6]{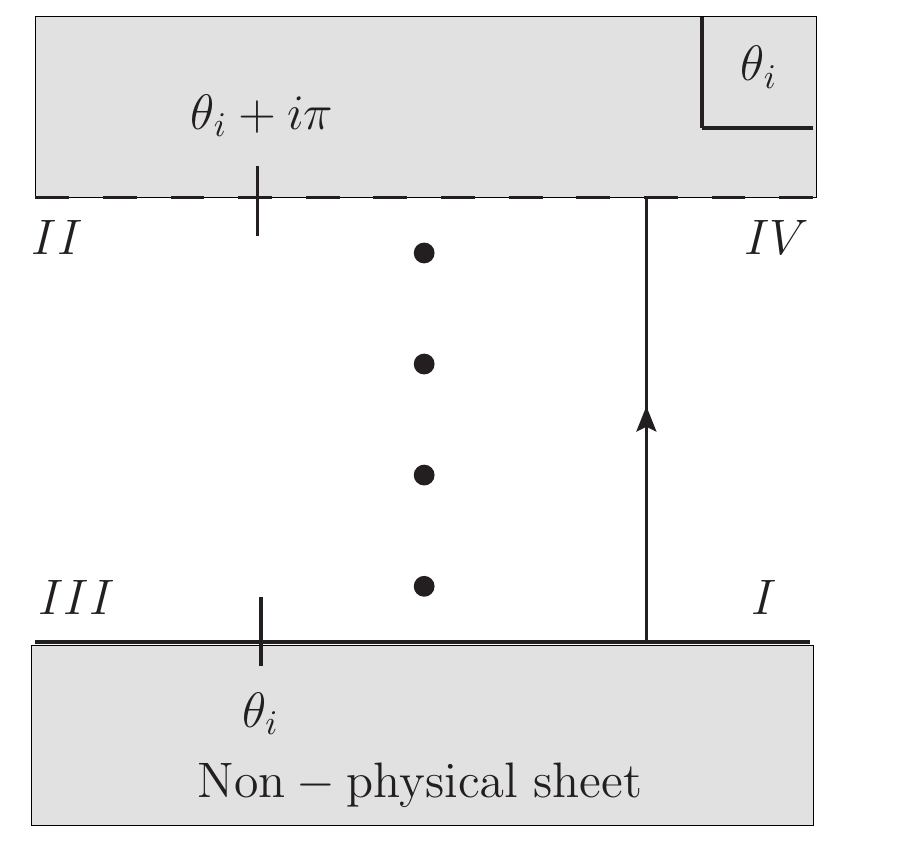}
\end{center}
\caption{Crossing for a Lorentz Invariant form factor in terms of one rapidity variable.}
\label{fig:Crossing_LI_ff_rapidity}
\end{figure}

As for the S-matrix, it is convenient to introduce the rapidity variables, $\theta_i$, and their differences, $\theta_{ij}=\theta_i-\theta_j$, in terms of which the auxiliary functions are defined by 
\<
f^{\cal O}_{\underline i}(\theta_1, \dots, \theta_n)=F^{\cal O}_{\underline i}(|\theta_{ij}|)~, ~~~{\rm for}~~\theta_1>\dots>\theta_n~,
\>
and for other configurations by analytical continuation. In terms of the rapidities with $\theta_i>\theta_j$ crossing corresponds to $\theta_{ij}\leftrightarrow i\pi -\theta_{ij}$. However, as there is the no cut in the t-channel, this is equivalent to $\theta_{ij}\leftrightarrow i\pi +\theta_{ij}$ so that this is also equivalent to $\theta_i\rightarrow i \pi +\theta_i$ or $\theta_j \rightarrow -i \pi +\theta_j$, see \figref{fig:Crossing_LI_ff_rapidity}.

\subsection{String Worldsheet Theory}

We now turn to the AdS$_5 \times$ S$^5$ worldsheet S-matrix. This is by now reasonably standard material and  we will follow closely the reviews  
\cite{Arutyunov:2009ga,Beisert:2010jr}. In the light-cone gauge fixed theory the fundamental on-shell excitations are the bosonic fields $\vecarrow{Y} = (Y_{i'=1,\ldots,4})$ and $\vecarrow{Z} = (Z_{i=5,\ldots,8})$, respectively corresponding to transverse excitations in the  S$^5$ and AdS$_5$ spaces, respectively, and the fermions, $\psi$, a Majorana-Weyl $\grSO(8)$ spinors of positive chirality. The symmetry preserved by the vacuum is  $\alg{psu}(2|2)^2\ltimes \mathbb{R}^3$ and so each particle, also called a magnon,
is characterized by an  $\alg{psu}(2|2)^2$ index, $i=(A,\dot A)$ where $A, \dot A=1, \dots, 4$. It useful to replace the momenta, $p$, of the 
massive excitations with two variables, $x^\pm$, such that 
\<
\frac{x^+}{x^-}=e^{i p}~,~~~{\rm and}~~~ x^++\frac{1}{x^+}-x^--\frac{1}{x^-}=\frac{2i }{g}
\> 
where $g$ is the coupling (related to the string coupling by $g^2=\tfrac{\lambda}{4\pi^2}$).\footnote{The fundamental representation, and tensor products thereof, also depend in the central charge parameter $\zeta$, see e.g. \cite{Beisert:2006qh}.} The dispersion relation is given by
\<
 E^2=1+4g^2 \sin^2 \frac{p}{2}~, ~~~{\rm or}~~~ E=\frac{i g}{2} \Big[x^- -\frac{1}{x^-}-x^+ +\frac{1}{x^+}\Big]~.
\>
It is also useful to define a parameter $u$, 
\<
\label{eq:uparam}
u=\frac{1}{2}\Big[x^+ +\frac{1}{x^+}+ x^- +\frac{1}{x^-}\Big]~.
\>
Below we will mostly focus on an $\alg{su}(2)$ sector of the theory involving a single complex bosonic field $Y$. The scattering
of two such $Y$-excitations with parameters $x_1^\pm$ and $x_2^\pm$ is described by the S-matrix
\<
\label{eq:exact_Smat}
\Smat=\sigma(x_1^\pm,x_2^\pm)^2~ \frac{u(x_1^\pm)-u(x_2^\pm)+\tfrac{i}{g}}{u(x_1^\pm)-u(x_2^\pm)-\tfrac{i}{g}}~,
\>
where $\sigma(x_1^\pm,x_2^\pm)$ is the so-called dressing phase, first determined by \cite{Beisert:2006ib, Beisert:2006ez}, and the remaining
term is the BDS S-matrix \cite{Beisert:2004hm}. 

The magnon dispersion relation is naturally uniformized in terms of Jacobi elliptic functions \cite{Janik:2006dc},
\<
p=2~{\rm am}~ z~, ~~~\sin \frac{p}{2}={\rm sn}(z,k)~, ~~~ E={\rm dn}(z,k)~,
\>
where $k=-4 g^2<0$. These expressions are naturally defined on the torus with real period $2\omega_1=4 K(k)$ and imaginary period $2\omega_2=4iK(1-k)-4K(k)$ with $K(k)$ the elliptic integral of the first kind. The dispersion relation is invariant under shifts of $z$, the analogue of the relativistic rapidity parameter, by $2\omega_1$ and $2\omega_2$. The real $z$-axis can be taken to be the physical region as for these values the energy is positive and the momentum real. The $x^\pm$ parameters are given by
\<
x^\pm=\frac{1}{2g}\left(\frac{{\rm cn} (z,k)}{{\rm sn}(z,k)}\pm i \right)(1+{\rm dn}( z,k))~,
\>
such that for real values of $z$ we have $|x^\pm|>1$ and ${\rm Im}(x^+)>0$ while ${\rm Im}(x^-)<0$. 

The crossing transformation corresponds to shifting $z$ by half the imaginary period, under which the positive branch of the dispersion transforms into the negative one, i.e.
\<
E(z)\rightarrow E(z+\omega_2)=-E(z)~, ~~~{\rm and}~~~p(z)\rightarrow p(z+\omega_2)=-p(z)~.
\>
Also, under crossing the parameters $x^\pm$ are transformed as: $x^\pm \rightarrow \tfrac{1}{x^\pm}$. The crossing transformation implies for the two-body S-matrix that
\<
S_{i_1 i_2}^{j_1j_2}(z_1+\omega_2,z_2)~(C_1^{-1})_{j_1j'_1}~S^{j'_1i'_2}_{j''_1 j_2}(z_1,z_2)~(C_1)^{j''_1i'_1}=\delta_{i_1}^{i'_1}\delta^{i'_2}_{i_2}
\>
The matrix $C_1$ is the charge conjugation matrix acting on the particle with momentum $p_1$. 

It is interesting to consider the limits $g\rightarrow \infty$ and $g\rightarrow 0$, i.e.\ strong and weak coupling. Taking $g\rightarrow \infty$ while rescaling $z\rightarrow \tfrac{z}{2 g}$ so that $p\rightarrow \tfrac{p}{g}$, the dispersion relation becomes relativistic and $z$ becomes the usual rapidity variable, $p=\sinh z$. The (rescaled) half-periods become
\<
\omega_1\rightarrow {2 \log g}~, \qquad \text{and} \qquad \omega_2\rightarrow i\pi~.
\>
From which we see that the torus degenerates into the infinite strip with $-\pi\leq {\rm Im}(z)\leq \pi$ i.e. twice the usual relativistic strip. As $g\rightarrow 0$, which corresponds the one-loop gauge theory, the half-periods become $\omega_1\rightarrow \pi$ and $\omega_2\rightarrow 2i \log g$ so that the crossing transformation becomes infinitely large. 
Another limit which will be important below is the so-called near-flat, or Maldacena-Swanson, limit \cite{Maldacena:2006rv}. This limit corresponds to focusing on the sector of worldsheet excitations with light-cone momenta, $p_\pm=\half(E\pm p)$, which scale as $p_\pm \sim g^{\mp1/2}$. In this limit
\<
p_-=e^{-z}~, ~~~{\rm and}~~~ p_+=e^z\left(1-\frac{e^{-4z}}{48}\right)~,
\>
which corresponds to the correct limit of the exact dispersion relation, see \cite{Maldacena:2006rv, Klose:2007rz}.

Finally, as can be seen from the pole structure of the S-matrix, specifically the BDS part, the theory possesses additional bound states of $n$ magnons. The bound states are most easily described in terms 
of the $n$ $u$-parameters, 
\<
\label{eq:bound_state}
u_k=u+ i \frac{(n-2k+1)}{g}~,~~~{\rm with}~~~k=1,\dots,n~. 
\>
These $n$-magnon states have the same dispersion relation as the single magnon but now with
\<
x^++\frac{1}{x^+}-x^--\frac{1}{x^-}=\frac{2n i }{g}~.
\>
These magnons can also be described in terms of the generalised rapidity parameters, $z$. For example, 
in the two-particle bound state there are two families of BPS magnons. Those with momenta below the critical 
value, $|p|<p_{\rm cr}$,
\footnote{The critical momenta is given by the formula 
$\sin^2 \frac{p_{\rm cr} }{2} =\frac{1}{2g^2} \left(\sqrt{1+4 g^2}-1\right)$.} for which the constituent particles have rapidities satisfying $z_1^\ast=z_2$
and those above criticality for which $z_1^\ast=-z_2+\tfrac{\omega_1}{2}+\tfrac{\omega_2}{2}$. The bound states then lie on a closed curve on the rapidity torus, see \cite{Arutyunov:2007tc} for a thorough description. 
In addition, the dressing phase multiplying the BDS part contributes double poles to the S-matrix. These are due to the exchange of pairs of BPS magnons \cite{Dorey:2007xn}.

\subsection{Form Factor Axioms}
\label{sec:axioms}

Our proposed set of consistency properties for the worldsheet form factors are simple generalizations of the more familiar 
axioms, as described for example by Smirnov \cite{Smirnov:1992vz}, in relativistic integrable theories. 
We will thus consider the form factors defined by
\<
 f_{i_1,\dots, i_n}^{\mathcal{O}}(z_1,\dots,z_n)=\bra{\Omega}{\cal O}\ket{p(z_1),\dots,p(z_n)}^{\rm (in)}_{i_1,\dots,i_n}~,
\>
as meromorphic functions of the torus parameters, $z_\alpha$, $\alpha=1,\dots, n$, of each external particle with the following properties.
\begin{itemize}
\item{Permutation:}
\<
\label{eqn:permutation-axiom}
 f_{\dots, i'_{l+1}, i'_{l}, \dots}(\dots, z_{l+1}, z_{l}, \dots) = f_{\dots, i_{l}, i_{l+1}, \dots}(\dots, z_l, z_{l+1}, \dots) 
\Smat^{i_{l} i_{l+1}}_{ i'_l i'_{l+1} }(z_l, z_{l+1})
\>
\item{Periodicity:}
\<
\label{eqn:periodicity-axiom}
f_{i_1,i_2,\dots, i_n}(z_1+\omega_2, z_2, \dots,z_n) = f_{i_2,\dots,i_n,i_1}(z_{2},\dots,z_n,z_1-\omega_2)
\>
\item{One-particle poles:} 
The form factors have poles in each subchannel corresponding to one-particle intermediate states goings on-shell, e.g., when $\vec{p}_{12}=\vec{p}(z_1)+\vec{p}(z_2)=0$
\<
\label{eq:one_particle_pole_axiom}
\Res_{\vec{p}_{12}=0}~ f_{i_1,\dots, i_n}(z_1,z_2,z_3,\dots, z_n) \eq 2 i C_{i_1i'_2} f_{i'_3,\dots,i'_n}(z_3,\dots,z_n) \nl \hspace{-10mm}
 \times\Big[\delta^{i'_2}_{i_2} \cdots \delta^{i'_n}_{i_n}-
 \Smat^{i'_2~~i'_n}_{j_{n-3} i_n}(z_n,z_2)\dots \Smat^{j_1 i'_3}_{i_2i_3}(z_3,z_2)\Big] \; .
\>
where $C_{i_1i_2}$ is the charge conjugation matrix introduced in the previous section. 
\item{Bound state poles:} 
As there are bound states in the worldsheet theory, the form factors will have additional poles, the residues of which are given by form factors with such bound states as external particles. Being somewhat schematic, and for simplicity considering a two-particle bound state in a rank one subsector of the full theory, if there is a pole in the (scalar) S-matrix at values of $z_1$ and $z_2$, that is $z'_1$ and $z'_2$ such that $u(z'_2)-u(z'_1)=\tfrac{2i}{g}$, where $u$ is the rapidity parameter
defined in \eqref{eq:uparam}, with the residue 
\<
\label{eqn:boundstate-axiom}
\Res_{z'_1,z_2'} \Smat_{12}(z_1, z_2)= R_{(12)}
\>
then the form factor will also have a pole at the values $z'_1$ and $z'_2$ and  
\<
\Res_{z'_1,z_2'} ~f(z_1,z_2,z_3,\dots, z_n)&=& \sqrt{2 i R_{(12)}} f(z_{12},z_3,\dots, z_n)~,
\>
where by the notation $z_{12}$ we denote the generalised rapidity parameter for the 
bound state. For example, a two-particle bound state occurs below the critical momentum, i.e. with both particles having equal real momenta, for $u_{1,2}=u_0 \pm\tfrac{i}{g}$, $u_0\in \mathbb{R}$. This  corresponds to $z_1^\ast=z_2$ and $z_2$ restricted to a curve such that ${\rm Im}(x_2^+)=0$ (see \cite{Arutyunov:2007tc}
for a more complete description of worldsheet bound states).
\end{itemize}

\section{Perturbative Computation of Form Factors}
\label{sec:perturbative-strong}

We will now perturbatively check the above axioms for the string worldsheet theory in various limits. At large effective string tension, $\sqrt{\lambda}\gg1$, the light-cone worldsheet theory is simply a (slightly complicated) two-dimensional theory of interacting bosons and fermions. For simplicity we restrict to an $\grSU(2)$ subsector of the theory, that is, we restrict to external states involving a single complex scalar
\<
  Y = \frac{1}{\sqrt{2}}(Y_1+i Y_2) \; .
\>
If there are only $Y$-particles in the external state this is a closed $\grSU(2)$ sector and so the mixing problem of states is greatly reduced. However we will also allow $\bar{Y}$-particles, which under crossing are $Y$-particles in the out-state.
The calculation of the form factors is then standard: for 
a  scalar field $Y(\vec{x})$ of mass $m$ and asymptotic particles with on-shell incoming momenta $p_i$, $i=1,\dots, n$ and outgoing momenta
$p'_j$, $j=1,\dots, m$, the LSZ formula relating worldsheet correlation functions to the connected component of asymptotic matrix elements is, in our notations, 
\<
{}^{\rm (out)} \bra{ p'_m,\dots,p'_1 }{\mathcal O}(\vec{x})\ket{p_1\dots p_n}^{\rm (in)}_{\rm connected}&=&\nn \\
& &\kern-140pt\lim_{\substack {\vec{p}_{i,0}\rightarrow E_i\\ \vec{p}'_{j,0}\rightarrow E'_j}}
\prod_{i=1}^n \int d^2 x_i e^{-i \vec{p}_i\cdot \vec{x}_i} (\sqrt{Z_i}\Delta_i)^{-1}
\prod_{j=1}^m\int d^2 y_j e^{i \vec{p}'_j \cdot \vec{y}_j} (\sqrt{Z'_j} \Delta'_j)^{-1}\nn \\
& &\kern-80pt\times \langle T\{\phi(\vec{x}_1)\dots \phi(\vec{x}_n){\cal O}(\vec{x})\phi(\vec{y}_1)\dots \phi(\vec{y}_m)\}\rangle
\>
where $Z_i$ and $Z'_j$ are the wave-function renormalisation factors and the inverse propagators, $\Delta_i^{-1}=-i (\vec{p}_i^2+m^2+i \varepsilon)$,  with two-momenta are taken on-shell: $p_0=\epsilon(p)=\sqrt{p^2+m^2}$, $p_1=p$. Thus we simply need to evaluate the connected, amputated Feynman diagrams following from the string action. 

\subsection{Perturbative Computation in Near-Plane-Wave Model}
\label{sec:perturbative-npw}

To find the appropriate vertices an obvious starting point is the near-plane-wave expansion of the full light-cone string action. This can be viewed as a fluctuation expansion about the large-$J$ BMN vacuum or equivalently as a large string tension expansion in the small momentum limit, e.g.\ see \cite{Berenstein:2002jq, Frolov:2002av, Callan:2003xr, Callan:2004uv,Frolov:2006cc}. After gauge fixing the action is
\<
S=\frac{\sqrt{\lambda}}{2\pi}\int d\tau \int^{\frac{L}{2}}_{-\frac{L}{2}}d\sigma ~{\cal L}~,
\>
where the length $L$ of the worldsheet is related to the vacuum angular momentum $J = \sqrt{\lambda} \mathcal{J}$ and the target space energy $E = \sqrt{\lambda} \mathcal{E}$ by
\< \label{eqn:gauge-parameter}
  \frac{L}{2\pi} = (1-a)\mathcal{J} + a\mathcal{E} \; .
\>
where $a$ is a parameter related to the specific light-cone gauge choice. The $Y$-part of the Langrangian density is given to quartic order in the fields by\footnote{For the derivatives $\partial \equiv (\partial_\tau,\partial_\sigma)$, we also use dot and prime notation, $\partial_\tau X=\dot X$ and  $\partial_\sigma X=\acute{X}$.}
\<
\label{eqn:agauge_pp_action}
  \Lagr = \partial Y \partial \bar{Y} - Y \bar{Y} + 2 Y \acute{Y} \bar{Y} \acute{\bar{Y}} + \frac{1-2a}{2} \Bigbrk{ (\partial Y)^2 (\partial \bar{Y})^2 -  Y^2 \bar{Y}^2 } \; .
\>
While the quartic part of the action is not Lorentz invariant, the quadratic part is and the implied index contractions are performed with the metric of signature $(+-)$. As for the perturbative calculation of the S-matrix \cite{Klose:2006zd}, in order to properly define asymptotic states it is necessary to take the decompactification limit $L\rightarrow \infty$. 

The simplest form factors are those for the fundamental field, $\mathcal{O}(\vec{x}) = Y(\vec{x})$, itself. Here we will consider the simplest form factors of this operator. The one-particle form factor with a single $Y$-particle of momentum $p_1$ in the external state\footnote{In this limit the rapidity torus has become an infinite strip and we will label the states and form factors by the particle momenta rather than the torus parameter.} is given by  
\<
  f(p) = \bra{0} Y \ket{p} = \sqrt{Z(p)} \; .
\>
At tree-level, the wave-function is simply given by the on-shell particle energy, $Z(p) = \tfrac{1}{2\epsilon}$. It may in fact be most useful to use the one-particle form factor to set the normalization of the operator by absorbing the wave-function factor thus setting this matrix element to be one. However, we will continue to consider the bare operators at this point. 

The tree-level three-particle form factor with one $\bar{Y}$-particle of momentum $p_1$ and two $Y$-particles of momenta $p_2$ and $p_3$ in the external state is
\<
\label{eqn:tree_ff_0to3}
f(\bar{p}_1,p_2,p_3) = - 2 \frac{ (p_2+p_3)^2 - (1-2a) (\vec{p}_1\cdot\vec{p}_{123} \,  \vec{p}_2\cdot\vec{p}_3 + 1) }{\sqrt{8\eps_1\eps_2\eps_3} \; (\vec{p}_{123}^2 - 1) } \; ,
\>
where we have introduced the notation $\vec{p}_{ij\ldots} = \vec{p}_i + \vec{p}_j + \ldots$.
At this order most of the properties of the form factors outlined in \secref{sec:axioms} are trivial. Specifically both the permutation \eqref{eqn:permutation-axiom} and periodicity \eqref{eqn:periodicity-axiom} axioms hold with the S-matrix being the identity. Moreover, as the magnon bound states are not observable as small perturbations about the vacuum there are no additional bound state poles. This can be easily seen by expanding the S-matrix and noting that at leading order for the scattering of two particles with momenta $p$ and $p'$ there is no pole except at  $p =p'$. The reason for this can be seen by examining \eqref{eq:bound_state} where, in the limit of large $g$ while keeping $u$ fixed, there is no pole except at $u_1=u_2$. 

The remaining property is that of factorization, or the one-particle pole axiom \eqref{eq:one_particle_pole_axiom}. Let us consider the case where $\vec{p}_1+\vec{p}_2\rightarrow 0$ which puts in the propagator on-shell and gives rise to a pole, the residue of which should be
\< \label{eqn:npw-ff3-res}
  \Res_{\vec{p}_{12}=0} f(\bar p_1, p_2, p_3)=2 i C_{\bar Y Y} f(p_3)(1-\Smat_{YY}(p_3,p_2)) \; ,
\>
where $\Smat_{YY}(p_2,p_3)$ is scattering amplitude of two $Y$-particles and $C_{\bar Y Y}$ is the matrix element of the charge conjugation matrix relating $Y$- and $\bar{Y}$-particles. In order to satisfy $\vec{p}_1+\vec{p}_2\rightarrow 0$ we will analytically continue to the the crossed region $\vec{p_1}\rightarrow -\vec{p_1}$ and then take $p_1\rightarrow p_2$. In this limit the residue of the propagator is $\epsilon_2/[2(\epsilon_2 p_3-\epsilon_3 p_2)]$ which combines with the wave-function factors, $(4 \epsilon_1\epsilon_2)^{-1/2} \rightarrow  i/2 \epsilon_2$, and the numerator to reproduce the leading interaction part of two-particle S-matrix, that is the near-plane-wave T-matrix\footnote{There is an additional factor of the worldsheet coupling $\tfrac{2\pi}{\sqrt{\lambda}}$ in front of both the T-matrix and the three-particle form factor that we have not explicitly included.}
\cite{Klose:2006zd},
\<
  \Tmat_{YY}(p_3,p_2)=\frac{i}{2 (\epsilon_2 p_3-\epsilon_3 p_2)} \Bigbrk{ (p_2+p_3)^2+(1-2a) (\epsilon_2 p_3-\epsilon_3 p_2)^2 } \; .
\> 
The remaining wave-function factor $1/\sqrt{2\epsilon_3}$ simply gives the one-particle form factor $f(p_3)$. Finally, the charge conjugation in this sector is a constant which we take to be $C_{\bar Y Y}=i/2$.

While the near-plane-wave action is a natural starting point for perturbative consideration of the form factors it is technically difficult to go beyond tree-level where the analytic structure, and so the form factor axioms, are essentially trivial. We will thus now turn to the so-called near-flat limit \cite{Maldacena:2006rv}.

\subsection{Perturbative Computation in Near-Flat-Space Model}

The near-flat-space limit \cite{Maldacena:2006rv} of the string sigma model on $\AdS_5\times\Sphere^5$ is a large radius limit ($R^2\sim\sqrt{\lambda} \gg 1$) in combination with a boost of the worldsheet coordinates with parameter $\lambda^{1/4}$. This does not reduce the number of degrees of freedom compared to the plane-wave model, but it significantly simplifies their interactions by enhancing the derivative couplings for left-movers, $\partial_- \sim \lambda^{1/4}$, and suppressing them for right-movers, $\partial_+ \sim \lambda^{-1/4}$, where the light-cone derivatives are $\partial_\pm = \half (\partial_\tau \pm \partial_\sigma)$.

The resulting near-flat-space Lagrangian can be written as \cite{Klose:2007wq,Klose:2007rz}
\< \label{eqn:MS-action}
 \mathcal{L} \eq
  \tfrac{1}{2}(\partial \vecarrow{Y})^2-\tfrac{1}{2}\,\vecarrow{Y}^2
 +\tfrac{1}{2}(\partial \vecarrow{Z})^2-\tfrac{1}{2}\,\vecarrow{Z}^2
 +\tfrac{i}{2}\psi \tfrac{\partial^2+1}{\partial_-}\,\psi \nl[1mm]
 + \gamma\,(\vecarrow{Y}^2-\vecarrow{Z}^2)\bigbrk{(\partial_- \vecarrow{Y})^2+(\partial_- \vecarrow{Z})^2}
 +i\gamma\,(\vecarrow{Y}^2-\vecarrow{Z}^2)\psi\partial_-\psi \nl[1mm]
 +i\gamma\,\psi\bigbrk{\partial_- Y_{i'} \Gamma_{i'} + \partial_- Z_i \Gamma_i}
           \bigbrk{Y_{j'} \Gamma_{j'} - Z_j \Gamma_j}\psi \nl[1mm]
 -\tfrac{\gamma}{24}\bigbrk{\psi\Gamma_{i'j'}\psi\,\psi\Gamma_{i'j'}\psi
                      -\psi\Gamma_{ij}  \psi\,\psi\Gamma_{ij}  \psi} \; .
\>
The usual prefactor $\sqrt{\lambda}/(2\pi)$ of the string Lagrangian has been scaled away and is now present as $\gamma = \pi/\sqrt{\lambda}$ in front of the interaction terms. The bosonic fields $\vecarrow{Y} $ and $\vecarrow{Z} $ are the same transverse excitations as in the near-plane-wave limit and the eight fermionic degrees of freedom are
also described by an $\grSO(8)$ Majorana-Weyl spinor $\psi$.

Because the interaction terms contain $\partial_-$-derivatives but are free from $\partial_+$-derivatives, it is convenient to quantize the model with the light-cone coordinate $\sigma^+$ considered as time. Thus, the mode expansions of the fields are
\begin{align}
Y_{i'}(\vec{x}) & = \int\frac{dp_-}{2\pi} \frac{1}{\sqrt{2p_-}} \:
                    \Bigsbrk{ a_{i'}(p_-)      \, e^{-i\vec{p}\cdot\vec{x}}
                            + a^\dag_{i'}(p_-) \, e^{+i\vec{p}\cdot\vec{x}} } \; , \\
Z_i(\vec{x})    & = \int\frac{dp_-}{2\pi} \frac{1}{\sqrt{2p_-}} \:
                    \Bigsbrk{ a_i(p_-)      \, e^{-i\vec{p}\cdot\vec{x}}
                            + a^\dag_i(p_-) \, e^{+i\vec{p}\cdot\vec{x}} } \; , \\
\psi(\vec{x})   & = \int\frac{dp_-}{2\pi} \frac{1}{\sqrt{2}} \:
                    \Bigsbrk{ b(p_-)        \, e^{-i\vec{p}\cdot\vec{x}}
                            + b^\dag(p_-)   \, e^{+i\vec{p}\cdot\vec{x}} } \; ,
\end{align}
from which we read off the tree-level wave-functions, $Z_Y = Z_Z = 1/(2p_-)$ and $Z_\psi = 1/2$. There are corrections to the wave-functions starting at two loops \cite{Klose:2007rz}, but we will not need them here. The free bosonic and fermionic propagators are
\<
  \frac{i}{\vec{p}^2 - 1} \comma \frac{ip_-}{\vec{p}^2 - 1} \; ,
\>
and the free dispersion relation is $2 p_+ =\tfrac{1}{2 p_-}$.

Due to this worldsheet light-cone quantization, the component $p_+$ has to be interpreted as the energy of the particle and $p_-$ as its momentum. This also implies that the form factor axioms of \secref{sec:axioms} apply with all $p$'s replaced by $p_-$'s. In order to avoid having to write too many $\pm$-subscripts, we introduce the notation
\<
  p_+ \equiv \xi \qquad \text{and} \qquad p_- \equiv \eta \; .
\>

\subsubsection{One-field operator}

\begin{figure}
\begin{center}
\begin{tabular}{cc}
\subfigure[Tree level]{\label{fig:F3-feynman-tree}\includegraphics[height=20mm]{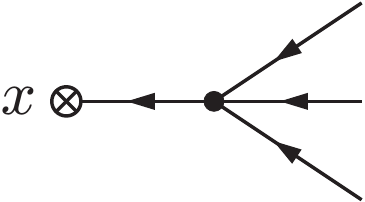}} \hspace{15mm} &
\subfigure[One loop]{\label{fig:F3-feynman-one}\includegraphics[height=20mm]{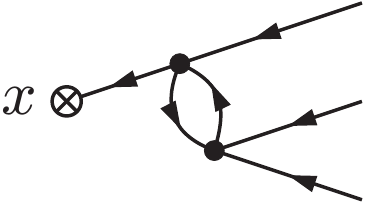}}
\end{tabular}
\end{center}
\caption{\textbf{Feynman diagrams for three-particle form factor.} There are two more one-loop diagrams which are obtained from this one here by permuting the external legs. Depending on which of the external legs corresponds to the anti-particle, not all of those three diagrams give a non-zero contribution to the form factor.}
\label{fig:F3-feynman}
\end{figure}

We begin by computing form factors for the fundamental field, $\mathcal{O}_1(\vec{x}) = Y(\vec{x})$. The one-particle form factor is again given by the wave-function
\< \label{eqn:ff1}
  f(\eta) = \sqrt{Z(\eta)} \; ,
\>
which is known to two loops \cite{Klose:2007rz}. In order to check the form factor axioms, however, we need to consider more than one external particle. Due to charge conservation, the next simplest form factor is the one for three external particles, one $\bar{Y}$, which we take to be the particle with $\eta_1$, and two $Y$'s with 
$\eta_2$ and $\eta_3$ . We compute this form factor perturbatively to one-loop order. The relevant Feynman diagrams are drawn in \figref{fig:F3-feynman}.
At tree-level we have, in a hopefully transparent notation, 
\< \label{eqn:ff3-tree}
  f^{(0)}(\bar{\eta}_1,\eta_2,\eta_3) \eq \frac{-\sqrt{2}\gamma} {\sqrt{\eta_1\eta_2\eta_3}}\, \frac{\eta_{23}^2}{\vec{p}_{123}^2-1} \; .
\>

At one-loop, the form factor is essentially a sum of bubble diagrams through which different combinations of the momenta of the external particles flow. The bubble is then connected by a propagator transferring the total momentum to the operator. The particles in the loop are not restricted to the $\grSU(2)$ sector, but can be any $Y_{i'}$, $Z_i$, or $\psi$. Summing up all those possibilities, we find\footnote{We have extended the multi-index notation to include differences: $\eta_{i..\bar{j}..} = \eta_i+ \ldots - \eta_k + \ldots$. All summands that come with a minus sign are dressed with a bar. This bar in completely unrelated to the bar in $\bar{\eta}_i$ which indicates that the particle that carries the momentum $\eta_i$ is a conjugate particle, here $\bar{Y}$.}
\< \label{eqn:ff3-one}
f^{(1)}(\bar{\eta}_1,\eta_2,\eta_3) \eq \frac{ \sqrt{8} i \gamma^2}{\sqrt{\eta_1\eta_2\eta_3}(\vec{p}_{123}^2-1)}
\biggsbrk{
  \eta_{23}^2 \Bigbrk{ \eta_{23}^2 B(\eta_{23}) + \eta_{1\bar{2}}^2 B(\eta_{12}) + \eta_{1\bar{3}}^2 B(\eta_{13}) } \nl[1mm] \hspace{40mm}
  + 4 \, (\eta_2^2 - \eta_3^2 - 2 \eta_1 \eta_3 ) \, \eta_1 \eta_2 B(\eta_{12}) \nl[1mm] \hspace{40mm}
  + 4 \, (\eta_3^2 - \eta_2^2 - 2 \eta_1 \eta_2 ) \, \eta_1 \eta_3 B(\eta_{13})
} \; ,
\>
where $B(\eta_{ij})$ is the bubble integral evaluated for the sum of two on-shell momenta, $\vec{p}_i+\vec{p}_j$, generally defined as
\< \label{eqn:bubble}
  B(\vec{p}) = \int \frac{d^2k}{(2\pi)^2} \frac{1}{[\vec{k}^2 - 1 + i\eps][(\vec{p}-\vec{k})^2 - 1 + i\eps]} \; .
\>
After expressing the $\xi$-components of the momenta by the $\eta$-components using the mass-shell condition, the bubble integral depends on $\eta_i$ and $\eta_j$ separately and not just on their sum. Moreover, determined by the $i\eps$-prescription, it evaluates to different expressions in different kinematical regions. If the loop integral is evaluated using the residue theorem, then this effect can be traced back to the fact that poles move in and out of the integration contour depending on the signs $\eta_i$ and $\eta_j$ and also on their relative sign. The result is
\< \label{eqn:bubble-cases}
  B(\eta_1,\eta_2) = 
  \frac{i}{2\pi} \frac{\eta_1 \eta_2}{\eta_1^2 - \eta_2^2} \begin{cases}
  \ln\lrbrk{\frac{\eta_2}{\eta_1}} - i\pi & \mbox{for $0 < \eta_1 < \eta_2$ or $\eta_2 < \eta_1 < 0$} \; , \\[2mm]
  \ln\lrbrk{-\frac{\eta_2}{\eta_1}}  & \mbox{for $\eta_1 < 0 < \eta_2$ or $\eta_2 < 0 < \eta_1$} \; , \\[2mm]
  \ln\lrbrk{\frac{\eta_2}{\eta_1}} + i\pi & \mbox{for $\eta_1 < \eta_2 < 0$ or $0 < \eta_2 < \eta_1$} \; .
  \end{cases}
\>
Note, however, that despite its complicated appearance, this formula is nonetheless symmetric in $\eta_1$ and $\eta_2$, which we can make manifest by writing it as
\<
  B(\eta_1,\eta_2) =
  \frac{i}{2\pi} \frac{ \eta_1 \eta_2}{\eta_1^2 - \eta_2^2} \ln\lrabs{\frac{\eta_2}{\eta_1}} - 
  \frac{\eta_1 \eta_2}{4\bigbrk{\eta_1+\eta_2}\abs{\eta_1-\eta_2}} \lrbrk{\frac{\eta_1}{\abs{\eta_1}} + \frac{\eta_2}{\abs{\eta_2}}}
  \; .
\>
The bubble integral had to be symmetric, because there is nothing that distinguishes the two momenta. As a consequence, also the matrix element, \mbox{$\bra{\Omega}\mathcal{O} \, a_{\bar{Y}}^\dagger(\eta_1) a_Y^\dagger(\eta_2) a_Y^\dagger(\eta_3) \ket{\Omega}$}, is symmetric under $\eta_2\leftrightarrow\eta_3$ which is no more than consistent with the fact that the creation operators $a_Y^\dagger(\eta_2)$ and $a_Y^\dagger(\eta_3)$ commute. However, this matrix element is not the form factor that satisfies the axioms.

Recall that the form factor is actually defined, in \eqref{eqn:def-aux-ff}, for an in-scattering state, i.e.\ we should assume $\eta_1 > \eta_2 > \eta_3 > 0$ in the case at hand and work with the functional form of $f(\bar{\eta}_1,\eta_2,\eta_3)$ computed in this particular kinematical region. Then, we define $f$ outside of this region by analytic continuation, rather than by the result of the Feynman diagram computation. This now boils down to choosing one of the three forms of the bubble in \eqref{eqn:bubble-cases} and using it for all values of the $\eta$'s. To be precise, we replace $B(\eta_{12}) \to B(\eta_1,\eta_2)$, $B(\eta_{13}) \to B(\eta_1,\eta_3)$, and $B(\eta_{23}) \to B(\eta_2,\eta_3)$ in \eqref{eqn:ff3-one} and then always use the third line in \eqref{eqn:bubble-cases} no matter what the relative signs of the momenta are that we plug into $f$. This analytically continued function is what we will mean when we write $f(\bar{\eta}_1,\eta_2,\eta_3)$ in the following and this function is no longer symmetric in $\eta_2$ and $\eta_3$.

\paragraph{Permutation.} We will now see how the permutation property \eqref{eqn:permutation-axiom} comes about. Let us first consider the permutation of the two $Y$ particles, i.e.\ those with momenta $\eta_2$ and $\eta_3$. This is simpler than, say, the permutation of $\bar{Y}(\eta_1)$ and $Y(\eta_2)$, because two $Y$ particles cannot scatter into any other particle species and, therefore, there will be only one term on the right hand side of \eqref{eqn:permutation-axiom}.

Let us compute $\Delta f^{(1)} \equiv f^{(1)}(\bar{\eta}_1,\eta_3,\eta_2) - f^{(1)}(\bar{\eta}_1,\eta_2,\eta_3)$. The only non-symmetric term is the bubble integral $B(\eta_2,\eta_3)$, for which we have $B(\eta_3,\eta_2) - B(\eta_2,\eta_3) = \eta_2\eta_3/\bigbrk{\eta_2^2-\eta_3^2}$, and thus
\< \label{eqn:ff3-delta23}
  \Delta f^{(1)} = \frac{ \sqrt{8} i \gamma^2}{\sqrt{\eta_1\eta_2\eta_3}(\vec{p}_{123}^2-1)} \frac{\eta_{23}^4 \eta_2\eta_3}{\eta_2^2-\eta_3^2} = \frac{-\sqrt{2}\gamma} {\sqrt{\eta_1\eta_2\eta_3}}\, \frac{\eta_{23}^2}{\vec{p}_{123}^2-1} \times (-2i\gamma) \eta_2\eta_3 \frac{\eta_2 + \eta_3}{\eta_2 - \eta_3} \; .
\>
We have written $\Delta f^{(1)}$ in a form, where we can recognize it as the product of the tree-level form factor, $f^{(0)}(\bar{\eta}_1,\eta_2,\eta_3)$, and the tree-level (order $\gamma$) piece, $\Smat^{(0)}_{YY}(\eta_2,\eta_3)$, of the S-matrix for the scattering of two $Y$-particles (see \appref{sec:near-flat-S-mat})
\<
  \Smat_{YY}(\eta_1,\eta_2) = S_0(A+B)^2 = 1 - 2i\gamma \eta_1 \eta_2 \, \frac{\eta_1+\eta_2}{\eta_1-\eta_2} + \order(\gamma^2) \; .
\>
Thus, we have obtained
\<
  f^{(1)}(\bar{\eta}_1,\eta_3,\eta_2) = f^{(1)}(\bar{\eta}_1,\eta_2,\eta_3) + f^{(0)}(\bar{\eta}_1,\eta_2,\eta_3) \Smat^{(0)}_{YY}(\eta_2,\eta_3) \; .
\>
Adding $f^{(0)}(\bar{\eta}_1,\eta_3,\eta_2) = f^{(0)}(\bar{\eta}_1,\eta_2,\eta_3)$ to this equation, we have verified the permutation property 
\<
  f(\bar{\eta}_1,\eta_3,\eta_2) = f(\bar{\eta}_1,\eta_2,\eta_3) \Smat_{YY}(\eta_2,\eta_3)
\>
up to one-loop level.

\bigskip

Next, we briefly look at the slightly more complicated case of permuting $\bar{Y}(\eta_1)$ and $Y(\eta_2)$. The permuted (or analytically continued) form factor $f(\eta_2,\bar{\eta}_1,\eta_3) \equiv f_{Y\bar{Y}Y}(\eta_2,\eta_1,\eta_3)$ is predicted to be equal to the sum of form factors $f_{X_1X_2Y}(\eta_1,\eta_2,\eta_3)$, where $X_1$ and $X_2$ are particles into which $\bar{Y}$ and $Y$ can scatter, times the corresponding S-matrix elements. Expressing this statement using $\grSU(2|2)^2$ index notation (see \appref{sec:near-flat-S-mat}), we have
\< \label{eqn:ff3-permutation}
  f_{1\dot{1},2\dot{2},1\dot{1}}(\eta_2,\eta_1,\eta_3) = f_{A\dot{A},B\dot{B},1\dot{1}}(\eta_1,\eta_2,\eta_3) \, \Smat_{2\dot{2},1\dot{1}}^{A\dot{A},B\dot{B}}(\eta_1,\eta_2) \; .
\>
Picking out the terms of order $\gamma^2$, this becomes
\< \label{eqn:ff3-permutation-12}
  f^{(1)}_{1\dot{1},2\dot{2},1\dot{1}}(\eta_2,\eta_1,\eta_3) - f^{(1)}_{2\dot{2},1\dot{1},1\dot{1}}(\eta_1,\eta_2,\eta_3) = f^{(0)}_{A\dot{A},B\dot{B},1\dot{1}}(\eta_1,\eta_2,\eta_3) \, \Smat^{(0)}{}_{2\dot{2},1\dot{1}}^{A\dot{A},B\dot{B}}(\eta_1,\eta_2) \; ,
\>
where the second term on the left hand side originates from the trivial (order $\gamma^0$) part of the S-matrix. At tree-level (order $\gamma^1$), the state $\ket{Y_{2\dot{2}}Y_{1\dot{1}}}$ scatters into
\< \label{eqn:scattering-of-YbarY}
 &&
 (A^2-1) \ket{Y_{2\dot{2}}Y_{1\dot{1}}}
 + AB \bigbrk{ \ket{Y_{1\dot{2}}Y_{2\dot{1}}} + \ket{Y_{2\dot{1}}Y_{1\dot{2}}} } \nl[1mm] \hspace{4mm}
 + AC \bigbrk{ \ket{\Psi_{2\dot{4}}\Psi_{1\dot{3}}} - \ket{\Psi_{2\dot{3}}\Psi_{1\dot{4}}}
             + \ket{\Bsi_{4\dot{2}}\Bsi_{3\dot{1}}} - \ket{\Bsi_{3\dot{2}}\Bsi_{4\dot{1}}} }
\>
with the coefficients
\<
  A^2 - 1 \eq - 2i\gamma \, \frac{\eta_1 \, \eta_2 \, \eta_{1\bar{2}}}{\eta_{12}} + \order(\gamma^2) \; , \\
  AB      \eq - 4i\gamma \, \frac{\eta_1^2 \, \eta_2^2}{\eta_{12} \, \eta_{1\bar{2}}} + \order(\gamma^2) \; , \\
  AC      \eq   2i\gamma \, \frac{\eta_1^{3/2} \, \eta_2^{3/2}}{\eta_{12}} + \order(\gamma^2) \; .
\>
To compute the right hand side of \eqref{eqn:ff3-permutation-12}, we thus need to know the tree-level three-particle form factors with the particles in \eqref{eqn:scattering-of-YbarY} having momenta $\eta_1$ and $\eta_2$, respectively, and a third particle $Y$ of momentum $\eta_3$. These form factors are given by
\begin{align}
& f^{(0)}_{{2\dot{2}},{1\dot{1}},{1\dot{1}}} =
  \frac{-\sqrt{2}\gamma} {\sqrt{\eta_1\eta_2\eta_3}}\, \frac{\eta_{23}^2}{\vec{p}_{123}^2-1} \; , \\
& f^{(0)}_{{1\dot{2}},{2\dot{1}},{1\dot{1}}} =
  f^{(0)}_{{2\dot{1}},{1\dot{2}},{1\dot{1}}} =
  \frac{-\sqrt{2}\gamma}{\sqrt{\eta_1\eta_2\eta_3}} \, \frac{\eta_2\,\eta_1 - \eta_3\,\eta_{123}}{\vec{p}_{123}^2-1} \; , \\
& f^{(0)}_{{2\dot{4}},{1\dot{3}},{1\dot{1}}} =
- f^{(0)}_{{2\dot{3}},{1\dot{4}},{1\dot{1}}} =
  f^{(0)}_{{4\dot{2}},{3\dot{1}},{1\dot{1}}} =
- f^{(0)}_{{3\dot{2}},{4\dot{1}},{1\dot{1}}} =
  \frac{\sqrt{2}\gamma}{\sqrt{\eta_3}} \, \frac{\eta_{23}}{\vec{p}_{123}^2-1} \; ,
\end{align}
where the first one is nothing but \eqref{eqn:ff3-tree}.

Finally, we are in the position to verify \eqref{eqn:ff3-permutation-12}. Summing the products of the tree-level form factors and the S-matrix elements collected above, we find that the right hand side evaluates to
\< \label{eqn:ff3-delta12}
\frac{ \sqrt{8} i \gamma^2}{\sqrt{\eta_1\eta_2\eta_3}(\vec{p}_{123}^2-1)}
\biggsbrk{
  \eta_{23}^2 \eta_{1\bar{2}}^2 + 4 \, (\eta_2^2 - \eta_3^2 - 2 \eta_1 \eta_3 ) \, \eta_1 \eta_2
  } \frac{\eta_1\eta_2}{\eta_1^2-\eta_2^2} \; .
\>
We also see that this is equal to the left hand side by noting that the asymmetry of the expression \eqref{eqn:ff3-one} in $\eta_1\leftrightarrow\eta_2$ stems solely from the bubble $B(\eta_1,\eta_2)$. Using $B(\eta_2,\eta_1) - B(\eta_1,\eta_2) = \eta_1\eta_2/\bigbrk{\eta_1^2-\eta_2^2}$, we find that $\Delta f^{(1)} \equiv f^{(1)}(\eta_2,\bar{\eta}_1,\eta_2) - f^{(1)}(\bar{\eta}_1,\eta_2,\eta_3)$ is precisely given by \eqref{eqn:ff3-delta12}.

\paragraph{Periodicity.} For the current case, the periodicity property \eqref{eqn:periodicity-axiom} reads
\< \label{eqn:ff3-periodicity}
  f(\bar{\eta}_1 e^{2\pi i}, \eta_2, \eta_3) \eq f(\eta_2, \eta_3, \bar{\eta}_1) \; .
\>
Combining it with the permutation property, we can also write this equation as $f(\bar{\eta}_1 e^{2\pi i}, \eta_2, \eta_3) = f(\eta_1, \eta_2, \eta_3)\Smat(\eta_1,\eta_2)\Smat(\eta_1,\eta_3)$, where matrix indices have been suppressed. The relation \eqref{eqn:ff3-periodicity} at one-loop order is again due to a property of the bubble integral. Because of the way we defined the form factor above, the third line in \eqref{eqn:bubble-cases} applies and it follows that
\<
  B(\eta_1 e^{2\pi i},\eta_2) - B(\eta_1,\eta_2) = \frac{\eta_1\eta_2}{\eta_1^2-\eta_2^2} = B(\eta_2,\eta_1) - B(\eta_1,\eta_2) \; .
\>
We see that inserting $\eta_1 e^{2\pi i}$ in place of $\eta_1$ yields the same change of $f(\bar{\eta}_1, \eta_2, \eta_3)$ as changing the relative signs of $\eta_1$ and $\eta_2$, and of $\eta_1$ and $\eta_3$, keeping the relative sign of $\eta_2$ and $\eta_3$ fixed. In formulas, this is
\<
  f^{(1)}(\bar{\eta}_1 e^{2\pi i}, \eta_2, \eta_3) - f^{(1)}(\bar{\eta}_1, \eta_2, \eta_3) \eq f^{(1)}(\eta_2, \eta_3, \bar{\eta}_1) - f^{(1)}(\bar{\eta}_1, \eta_2, \eta_3) \; ,
\>
or simply \eqref{eqn:ff3-periodicity}.

\paragraph{One-particle pole.} The three-particle form factor has a pole where the anti-particle $\bar{Y}(\eta_1)$ cancels out one of the particles, say, $Y(\eta_2)$, i.e.\ when the sums of their energies and their momenta vanish, $\vec{p}_1+\vec{p}_2=0$, and the internal propagator goes on-shell. The residue of this pole is then related to the one-particle form factor for this operator and the appropriate S-matrix element by the same equation, \eqref{eqn:npw-ff3-res}, discussed in the near-plane-wave case in \secref{sec:perturbative-npw}. Now, however, we will be able to verify this axiom also at one-loop level.

The residue of the propagator in light-cone variables is $\Res_{\eta_1=-\eta_2} (\vec{p}_{123}^2 - 1)^{-1} = {\eta_2^2\eta_3}/{\eta_{23}\eta_{2\bar{3}}}$,
so that the residue of the tree-level and one-loop form factors are
\<
  \Res_{\eta_1=-\eta_2} f^{(0)}(\bar{\eta}_1,\eta_2,\eta_3) \!\!\!\eq\!\!\! \frac{2i\gamma}{\sqrt{2\eta_3}} \, \eta_2\eta_3 \, \frac{\eta_{23}}{\eta_{2\bar{3}}} \; , \\
  \Res_{\eta_1=-\eta_2} f^{(1)}(\bar{\eta}_1,\eta_2,\eta_3) \!\!\!\eq\!\!\! \frac{4\gamma^2}{\sqrt{2\eta_3}} \frac{\eta_2\eta_3}{\eta_{23}\eta_{2\bar{3}}} \Bigsbrk{
    \eta_{23}^4 B(\eta_2,\eta_3) + (\eta_2^4 + 6\eta_2^2\eta_3^2+\eta_3^4) B(-\eta_2,\eta_3) + 8 \eta_2^2 \eta_3^2 B(0) 
  }. \nln
\>
Using the explicit expression for the bubble integral, its analytic continuation, and its limit for vanishing momentum,
\<
  B(\eta_2,\eta_3) = \frac{i}{2\pi} \frac{\eta_2\eta_3}{\eta_2^2-\eta_3^2} \Bigbrk{ \ln\frac{\eta_3}{\eta_2} + i\pi }
  \; , \;\;
  B(-\eta_2,\eta_3) = - \frac{i}{2\pi} \frac{\eta_2\eta_3}{\eta_2^2-\eta_3^2} \ln\frac{\eta_3}{\eta_2}
  \; , \;\;
  B(0) = \frac{i}{4\pi}
  \; ,
\>
respectively, we can simplify the one-loop residue to
\<
  \Res_{\eta_1=-\eta_2} f^{(1)}(\bar{\eta}_1,\eta_2,\eta_3) = \frac{4\gamma^2}{\sqrt{2\eta_3}} \frac{\eta_2\eta_3}{\eta_{23}\eta_{2\bar{3}}} \biggsbrk{
    - \frac{1}{2} \eta_2\eta_3 \frac{\eta_{23}^3}{\eta_{2\bar{3}}}
    + \frac{2 i}{\pi} \eta_2^2 \eta_3^2
    + \frac{2 i}{\pi} \eta_2^2\eta_3^2 \frac{\eta_2^2+\eta_3^2}{\eta_{23}\eta_{2\bar{3}}} \ln\frac{\eta_3}{\eta_2}
  } \; .
\>

The residue is supposed to be equal to $2iC_{\bar{Y}Y} \, f(\eta_3) \, (1-\Smat_{YY}(\eta_3,\eta_2))$. Expanding the S-matrix for YY-scattering, $S_0(A+B)^2$ (see \appref{sec:near-flat-S-mat}) to order $\gamma^2$, we find
\< \label{eqn:ff3-pole-match}
1-\Smat_{YY}(\eta_3,\eta_2) = -2i\gamma \eta_2\eta_3 \frac{\eta_{23}}{\eta_{2\bar{3}}} + 2\gamma^2 \eta_2^2 \eta_3^2 \frac{\eta_{23}^2}{\eta_{2\bar{3}}^2} - \frac{8i\gamma^2}{\pi} \frac{\eta_2^3\eta_3^3}{\eta_{23}\eta_{2\bar{3}}} \lrbrk{ 1 + \frac{\eta_2^2+\eta_3^2}{\eta_{23}\eta_{2\bar{3}}} \ln\frac{\eta_3}{\eta_2}} \; .
\>
As this expression starts at order $\gamma$, we only need the zeroth order term of the form factor $f(\eta_3)$ which is given by $1/\sqrt{2\eta_3}$. Taking this factor as well as the factor $2iC_{\bar{Y}Y} = -1$ into account, we see that the four terms of \eqref{eqn:ff3-pole-match} match the tree-level and the three terms of the one-loop form factor perfectly.

\paragraph{Bound states.} The bound state poles of the exact form factors cannot be seen in perturbation theory about the trivial vacuum. As for the near-plane-wave expansion, examination of the explicit one-loop S-matrix reveals that there are no additional poles for complex momenta.

\subsubsection{Two-field operator}

We also compute a couple of one-loop form factors for the simplest composite operator, namely $\mathcal{O}_2(\vec{x}) = \Half :Y(\vec{x}) Y(\vec{x}):$, and verify that the axioms are satisfied.

\begin{figure}
\begin{center}
\subfigure[Tree level]{\includegraphics[height=20mm]{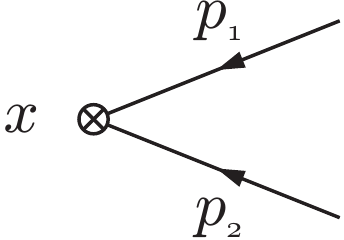}}
\hspace{30mm}
\subfigure[One loop]{\includegraphics[height=20mm]{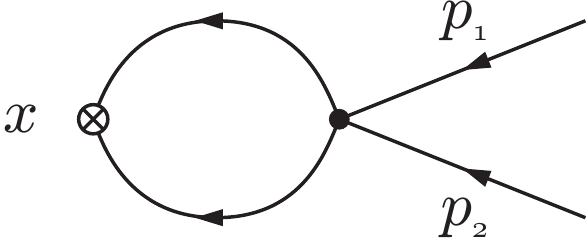}}
\end{center}
\caption{\textbf{Feynman diagrams for two-particle states.}}
\label{fig:F2-feynman}
\end{figure}

\paragraph{Two-particle form factor.} The Feynman diagrams for the form factor of $\mathcal{O}_2$ with $\ket{YY}$ as the external state are drawn in \figref{fig:F2-feynman}. At tree level, the result is just the product of the wave-functions
\< \label{eqn:ff2-tree}
  f^{(0)}(\eta_1,\eta_2) = \frac{1}{2\sqrt{\eta_1\eta_2}} \; ,
\>
and at one loop, we find the bubble integral with some numerator factors due to derivative couplings
\< \label{eqn:ff2-one}
  f^{(1)}(\eta_1,\eta_2) = \frac{- i \gamma}{\sqrt{\eta_1\eta_2}} \eta_{12}^2 B(\eta_{12}) \; .
\>
The verification of the axiom follows the same reasoning as above.

\begin{figure}
\begin{center}
\begin{tabular}{cc}
\subfigure[Tree level]{\label{fig:F4-feynman-tree}\includegraphics[height=20mm]{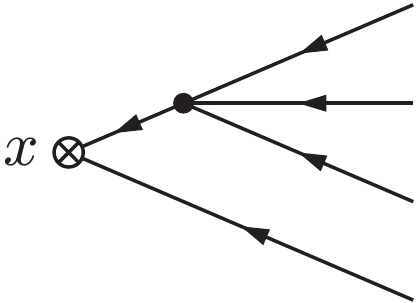}} \hspace{15mm} &
\begin{tabular}{p{38mm}p{38mm}}
\subfigure[One loop 1]{\label{fig:F4-feynman-one-1}\includegraphics[height=20mm]{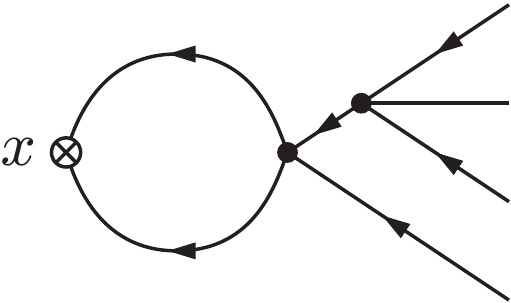}} &
\subfigure[One loop 2]{\label{fig:F4-feynman-one-2}\includegraphics[height=20mm]{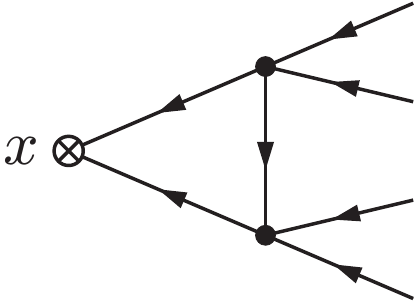}} \\[5mm]
\subfigure[One loop 3]{\label{fig:F4-feynman-one-3}\includegraphics[height=20mm]{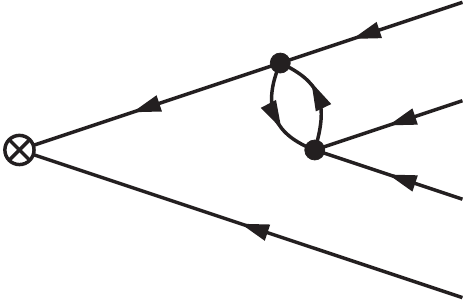}} &
\subfigure[One loop 4]{\label{fig:F4-feynman-one-4}\includegraphics[height=20mm]{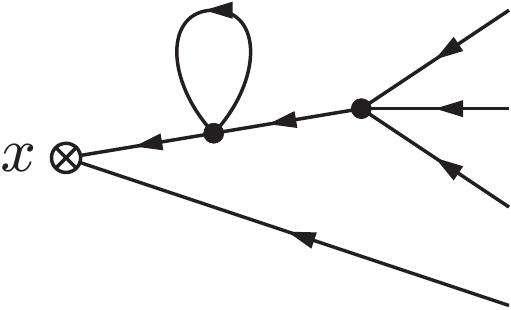}}
\end{tabular}
\end{tabular}
\end{center}
\caption{\textbf{Feynman diagrams for four-particle form factor.}}
\label{fig:F4-feynman}
\end{figure}

\paragraph{Four-particle form factor.} We have drawn the Feynman diagrams for the form factor of $\mathcal{O}_2$ with $\ket{\bar{Y}YYY}$ as the external state in \figref{fig:F4-feynman}. At tree-level, all diagrams contain one propagator that transfers the momentum of three of the in-coming particles to one of the fields in the operator while the remaining one only contributes its wave-function
\< \label{eqn:ff4-tree}
f^{(0)}(\bar{\eta}_1,\eta_2,\eta_3,\eta_4) \eq \frac{-\gamma}{\sqrt{\eta_1\eta_2\eta_3\eta_4}} \lrsbrk{
  \frac{\eta_{23}^2}{\vec{p}_{123}^2-m^2}
+ \frac{\eta_{24}^2}{\vec{p}_{124}^2-m^2}
+ \frac{\eta_{34}^2}{\vec{p}_{134}^2-m^2}
 } \; .
\>
At one loop, there are four types of diagrams which, taking the combinatorics into account, produce quite a large number of terms. As there are 6 fields involved---two in the operator and four in the external state---the Feynman diagram are the same as encountered in the 3-particle S-matrix computation in \cite{Puletti:2007hq}, except that here not all external lines are on-shell. The sum of the diagrams of type \figref{fig:F4-feynman-one-4} vanishes. This is the same cancellation as the one for the one-loop correction to the propagator. Diagrams of types \figref{fig:F4-feynman-one-1} and \figref{fig:F4-feynman-one-3}, directly yield products of propagators and bubbles. The diagrams of type \figref{fig:F4-feynman-one-2} are more complicated but can be reduced to propagators and bubbles using the cutting rule \cite{Kallen:1965}. The final result is
\< \label{eqn:ff4-one}
f^{(1)}(\bar{\eta}_1,\eta_2,\eta_3,\eta_4) \eq \frac{2i\gamma^2}{\sqrt{\eta_1\eta_2\eta_3\eta_4}}
\Biggl\{
    \biggsbrk{
      \frac{\eta_{23}^2}{\vec{p}_{123}^2-m^2}
    + \frac{\eta_{24}^2}{\vec{p}_{124}^2-m^2}
    + \frac{\eta_{34}^2}{\vec{p}_{134}^2-m^2}
  } \nl \hspace{27mm}
  \times \Bigsbrk{ \eta_{23}^2 B(\eta_{23}) + \eta_{24}^2 B(\eta_{24}) + \eta_{34}^2 B(\eta_{34}) \nl \hspace{30mm}
    + \eta_{1\bar{2}}^2 B(\eta_{12}) + \eta_{1\bar{3}}^2 B(\eta_{13}) + \eta_{1\bar{4}}^2 B(\eta_{14}) } \nl[1mm] \hspace{15mm}
  + 8 \biggsbrk{
      \frac{\eta_2^2 - \eta_3^2 - 2 \eta_1 \eta_3}{\vec{p}_{123}^2-m^2}
    + \frac{\eta_2^2 - \eta_4^2 - 2 \eta_1 \eta_4}{\vec{p}_{124}^2-m^2}
    } \eta_1 \eta_2 B(\eta_{12}) \nl[2mm] \hspace{15mm}
  + 8 \biggsbrk{
      \frac{\eta_3^2 - \eta_2^2 - 2 \eta_1 \eta_2}{\vec{p}_{134}^2-m^2}
    + \frac{\eta_3^2 - \eta_4^2 - 2 \eta_1 \eta_4}{\vec{p}_{124}^2-m^2}
    } \eta_1 \eta_3 B(\eta_{13}) \nl[2mm] \hspace{15mm}
  + 8 \biggsbrk{
      \frac{\eta_4^2 - \eta_2^2 - 2 \eta_1 \eta_2}{\vec{p}_{124}^2-m^2}
    + \frac{\eta_4^2 - \eta_3^2 - 2 \eta_1 \eta_3}{\vec{p}_{134}^2-m^2}
    } \eta_1 \eta_4 B(\eta_{14})
\Biggr\} \; . \hspace{5mm} \mbox{}
\>
And again, the discontinuities of the bubble integrals give rise to the characteristic form factor properties.

\section{Perturbative Computation at Weak Coupling}
\label{sec:perturbative-weak}

At weak 't\,Hooft coupling, the AdS/CFT dual of string energies  are the perturbative planar anomalous dimensions of gauge invariant operators, which can be calculated by means of an integrable spin-chain e.g. \cite{Minahan:2002ve}. Although the spin-chain model is still integrable, it is not only non-relativistic, but it is also a discrete system. It is thus 
interesting to check the form factor consistency conditions in this completely different regime and moreover to investigate whether
the relation between strings and spin-chains can be generically continued off-shell. The computation of integrable spin-chain form factors is well developed subject making use of techniques such as algebraic Bethe ansatz, the axiomatic q-deformed Knizhnik-Zamolochikov approach and quantum inverse scattering, see \cite{1993qism.book.korepin, Jimbo:1994qp, Kitanine1999647} or for more recent reviews \cite{Boos:2006ux, Kitanine:2005pu}. 
The use of algebraic Bethe ansatz methods for the calculation of spin-chain
form factors in the context of the AdS/CFT correspondence has previously been considered  \cite{Roiban:2004va} with the goal of calculating planar gauge theory
structure contexts. 
Building on this older work, see also \cite{Okuyama:2004bd, Alday:2005nd},
there has been recent progress in the problem making use of the integrable spin-chain description to calculate structure constants at weak coupling and indeed matching with strong coupling 
\cite{Escobedo:2010xs, Escobedo:2011xw, Georgiou:2011qk, Gromov:2011jh, Foda:2011rr, Bissi:2011ha, Gromov:2012vu, Ahn:2012uv, Foda:2012wf, Serban:2012dr, Kostov:2012yq, Gromov:2012uv,Bissi:2012vx}. However, with the comparison to the worldsheet computations of the previous section in mind, a more direct approach based on the coordinate Bethe ansatz is sufficient. 

We will in fact even find explicit agreement between the string and spin-chain calculations in the thermodynamic limit (see \secref{sec:matching-form-factors}). A priori, because of the different orders in which the limits on the gauge and string theory side are taken, this did not have to occur. It 
is very likely related to the agreement found in the computation of the one- and two-loop spectrum in the near-BMN limit \cite{Parnachev:2002kk,Callan:2003xr,Callan:2004uv} and, as in that case, a more general matching will require exact results going beyond the scope of this work.\footnote{The one- and two-loop agreement is presumably due to a currently unknown non-renormalization theorem. For the spectral problem, an argument circumventing the order of limits problem and hence explaining the matching was given in \cite{Harmark:2008gm}.}

Just as on the string theory side, we focus on the $\grSU(2)$ sector. In this sector, the one-loop dilatation generator of $\superN=4$ SYM is given by the Heisenberg $s=\half$ spin-chain Hamiltonian \cite{Minahan:2002ve}
\<
  H = \frac{\lambda}{8 \pi^2} \sum_{x=1}^L \lrbrk{ \unit - P }_{x,x+1}= \frac{\lambda}{16 \pi^2} \sum_{x=1}^L \lrbrk{1 - \vecsigma_{x}\cdot\vecsigma_{x+1} }  \; ,
\>
acting on a periodically identified spin-chain of length $L$. The ground state of zero energy is given by $\ket{0} = \ket{\uparrow\uparrow \cdots \uparrow}$. Spins are flipped by acting with the lowering operator $S_-$ and we denote the states in the ``coordinate basis'' by $\ket{x_1,x_2, \ldots} = S_{-,x_1} S_{-,x_2} \cdots \ket{0}$. The energy eigenstates are roughly the Fourier transformation of these states, rendering each flipped spin into a magnon with a momentum:
\< \label{eqn:Bethe-state}
  \ket{\psi(p_1,p_2,\ldots)} = \sum_{1\le x_1 < x_2 < \ldots \le L} \chi(p_1,p_2,\ldots)_{x_1,x_2,\ldots} \ket{x_1,x_2,\ldots} \; .
\>
It would be precisely a Fourier transformation if the wave-function $\chi(p_1,p_2,\ldots)_{x_1,x_2,\ldots}$ was given by $\prod_j e^{ip_jx_j}$ and the summation over the $x_j$ ranged from $1$ to $L$ without constraint. The actual eigenstates have the same structure, except that the portions of the wave functions that correspond to different orderings of the $x_j$ are normalized differently. For instance, for the two-magnon state, the wave functions is
\< \label{eqn:Bethe-wave-function}
  \chi(p_1,p_2)_{x_1,x_2} = e^{i(p_1 x_1+p_2 x_2)} + \Smat(p_2,p_1) \, e^{i(p_2 x_1+p_1 x_2)} \; ,
\>
where $\Smat(p_2,p_1)$ has the interpretation of the S-``matrix'' for the scattering of the magnons on the spin-chain. It is given by
\< \label{eqn:Heisenberg-Smatrix}
  \Smat(p_2,p_1) = - \frac{e^{i(p_1+p_2)} - 2 e^{ip_2} + 1}{e^{i(p_1+p_2)} - 2 e^{ip_1} + 1} \; ,
\>
and satisfies $\Smat(p_1,p_2) = 1/\Smat(p_2,p_1)$. For more than two magnons, the wave-function is a sum of as many terms as there are permutations of the momenta and each term is multiplied by a product of (two-magnon) S-matrices corresponding to the transpositions that are necessary to convert the ordered list of momenta into that particular permutation. It is not necessary but common to normalize the states such that the term in which $p_j$ goes with $x_j$ has \emph{no} factor besides the exponential.

The states \eqref{eqn:Bethe-state} with a wave-function of the form \eqref{eqn:Bethe-wave-function} are called Bethe states. They are energy eigenstates of the finite-$L$ spin-chain if and only if the momenta satisfy the Bethe equations
\< \label{eqn:Bethe-equations}
  e^{ip_kL} = \prod_{j\not=k} \Smat(p_k,p_j) \qquad \text{for all $k$} \; .
\>
Taking the product of these equations yields $\exp\lrsbrk{i(p_1+p_2+\ldots)L} = 1$.

\subsection{Properties of Spin-Chain Form Factors.}

In complete analogy to continuous models, a spin-chain form factor is the matrix element of an operator action on a specific site, or a few neighboring sites, taken between Bethe states
\<
  \bra{\psi(p'_1,\ldots)} \mathcal{O}_{x} \ket{\psi(p_1,\ldots)} \; ,
\>
where the subscript $x$ indicates the first site on which the operator acts. The $x$-dependence is again universal and can be found by using the shift operator $U(x)$, which shifts all spins by $x$ sites to the right. The actions of $\mathcal{O}_x$ and $\mathcal{O}_1$ are then related by $\mathcal{O}_x = U(x-1) \mathcal{O}_1 U(1-x)$. The external states, \eqref{eqn:Bethe-state}, are eigenstates of $U(x)$ with eigenvalue $e^{ip_{\mathrm{tot}}x}$. So, we can evaluate the $U$'s on the external states and find
\<
  \bra{\psi(p'_1,\ldots)} \mathcal{O}_x \ket{\psi(p_1,\ldots)} = e^{i(p_1+\dots-p'_1-\ldots)(x-1)} \bra{\psi(p'_1,\ldots)} \mathcal{O}_1 \ket{\psi(p_1,\ldots)} \; .
\>
This is the same $x$-dependence as in \eqref{eqn:ff-def-indices}. To obtain literally matching expressions, we should define the spin-chain form factor $F^{\mathcal{O}}$ for the operator acting on site $0\equiv L$, but this is less natural for the spin-chain.

\paragraph{Permutation.} In the spin-chain context, the permutation property \eqref{eqn:permutation-axiom} of the form factor, is a direct consequence of the properties of the Bethe states \eqref{eqn:Bethe-state}. The fact that the Bethe states acquire factors of the S-matrix when the momenta are permuted is inherited from the fact that the parts of the wave-function with different relative orderings of the momenta have different weights, as discussed above. For example, in the two-magnon case, we can compute
\<
  \chi(p_2,p_1)_{x_1,x_2} \eq e^{i(p_2 x_1+p_1 x_2)} + \Smat(p_1,p_2) \, e^{i(p_1 x_1+p_2 x_2)} \\
  \eq \Smat(p_1,p_2) \lrsbrk{ e^{i(p_1 x_1+p_2 x_2)} + \Smat(p_2,p_1) \, e^{i(p_2 x_1+p_1 x_2)} } = \Smat(p_1,p_2) \chi(p_1,p_2)_{x_1,x_2} \; , \nn
\>
where we used that $\Smat(p_1,p_2)$ and $\Smat(p_2,p_1)$ are inverses of each other. This implies the permutation property
\<
  \ket{\psi(p_2,p_1)} = \ket{\psi(p_1,p_2)} \, \Smat(p_1,p_2) \; .
\>
For more than two magnons, the state will acquire as many S-matrix factors as necessary to convert the two orderings into each other. This is precisely the permutation property \eqref{eqn:permutation-axiom}.

\paragraph{Periodicity and one-particle poles.} On the spin-chain side of the duality, where the 't Hooft coupling is small, one period of the rapidity torus becomes infinite and the periodicity property \eqref{eqn:periodicity-axiom} becomes invisible. Similarly it is not possible to go to the crossed channel such that the one-particle poles are not apparent
\footnote{One could of course consider the case where a magnon in the out-state has the same momenta as a magnon in the in-states. We will not consider
that here but see the explicit expressions below.}.

\paragraph{Bound States.} One of most interesting aspects of the spin-chain limit is
the abililty to study the parameter space where bound states exist so that the bound state condition \eqref{eqn:boundstate-axiom}
can be checked. 
Two-magnon bound states of the Heisenberg spin-chain are solutions of the Bethe equation of the form
\<
  \hat{p}_1 = \frac{p}{2} + i \Delta p \comma \hat{p}_2 = \frac{p}{2} - i \Delta p \; .
\>
Without loss of generality, we assume that $\Delta p>0$. Solutions of this kind are very easy to find analytically in the thermodynamic limit, $L\to\infty$, as they correspond to poles of the S-matrix $\Smat(p_2,p_1)$. The wave-function of the bound state has the same form as for real solutions, \eqref{eqn:Bethe-wave-function}, but acquires qualitatively different features. Firstly, the second term dominates over the first, and in the large $L$ limit, the wave-function should, in fact, be ``re-normalized'' such that it remains finite. This essentially amounts to taking the residue of the wave-function. Secondly, the dominant wave function,
\<
  e^{i(\hat{p}_2x_1+\hat{p}_1x_2)} = e^{-\Delta p(x_2-x_1)} e^{ip(x_1+x_2)/2} \; ,
\>
has an oscillatory part centered at the mean value of $x_1$ and $x_2$ and is damped for large distances $x_2 - x_1$, where we recall that the sum is over terms with $x_1<x_2$.

Given that the bound state wave-function is just the universal wave-function evaluated on the bound state momenta
\<
  \ket{\psi_B(\hat{p}_1,\hat{p}_2)} = \sum_{x_1<x_2} e^{i\hat{p}_2x_1+i\hat{p}_1x_2} \ket{x_1,x_2} \; ,
\>
the bound state axiom of the form factor \eqref{eqn:boundstate-axiom}, is again really a property of the Bethe states
\<
  \Res \ket{\psi(p_1,p_2)} = \ket{\psi_B(\hat{p}_1,\hat{p}_2)} \Res S(p_1,p_2) \; .
\>

\subsection{Examples}

In this subsection, we compute some form factors in the spin-chain setting. As always, the simplest form factor is the one for the fundamental operator, here $S_+$, and the one-particle state. It is given by
\<
  \bra{0} S_{+,x} \ket{\psi(p)} \eq e^{i p x} \; .
\>
Similarly, the $r$-magnon wave-functions can be extracted by a spin-operator of range $r$ as
\<
  \bra{0} S_{+,x} \ldots S_{+,x+r} \ket{\psi(p_1,\ldots,p_r)} \eq \chi(p_1,\ldots,p_r)_{x,\ldots,x+r} \; .
\>
This formula also holds if the spin operators do not act on adjacent sites. These are essentially all the form factors with the out-state being the vacuum chain. That is because of charge conservation the number of raising operators in the operator needs to agree with the number of magnons in the in-state for the result to be non-zero. If there were anti-particle excitations, then one could have $S_-$'s in the operator or $S_+$ excitations in the in-state.

The next simplest form factor is given by $\bra{\psi(p_1)} S_{+,x} \ket{\psi(p_2,p_3)}$ which we will compute now and then compare to the string theory result from \secref{sec:perturbative-npw} in \secref{sec:matching-form-factors}. By definition, it is given by
\<
  \bra{\psi(p_1)} S_{+,x} \ket{\psi(p_2,p_3)} = \sum_{\substack{1\leq x_1\leq L \\ 1\leq x_2<x_3\leq L}} \chi^*(p_1)_{x_1} \chi(p_2,p_3)_{x_2,x_3} \bra{0}S_{+,x_1}S_{+,x}S_{-,x_2}S_{-,x_3}\ket{0} \; .
\>
Using the fact that $x_2\neq x_3$, we have
\<
\bra{0}S_{+,x_1} \, S_{+,x} \, S_{-,x_2} \, S_{-,x_3}\ket{0} = \delta_{x,x_2}\delta_{x_1,x_3}+\delta_{x,x_3}\delta_{x_1,x_2}
\>
so that 
\<
  \bra{\psi(p_1)} S_{+,x} \ket{\psi(p_2,p_3)}
   = \sum_{x < x_3 \leq L} \chi^*(p_1)_{x_3} \chi(p_2,p_3)_{x,x_3}
   + \sum_{1 \leq x_2 < x} \chi^*(p_1)_{x_2} \chi(p_2,p_3)_{x_2,x} \; .
\>
Focusing on the $x=1$ case, the second sum does not contribute. Inserting the explicit form of the wave functions, we have
\< \label{eqn:ff-1to2-sums}
  \bra{\psi(p_1)} S_{+,1} \ket{\psi(p_2,p_3)} \eq
          e^{ip_2}                \sum_{1\leq x_3\leq L}e^{i (p_3-p_1) x_3}
        + e^{ip_3} \Smat(p_3,p_2) \sum_{1\leq x_3\leq L}e^{i (p_2-p_1) x_3} \nl[2mm]
        - e^{i(p_2+p_3-p_1)} \bigbrk{ 1 + \Smat(p_3,p_2) } \; ,
\>
where the last line compensates for the $x_3=1$ terms in the sums which were not present in the previous formula. The sums look like $\delta$-functions in the momenta and a rough interpretation of the three terms in a field theory language would be as follows. The first sum represent disconnected diagrams where particle 3 emerges as particle 1 in the out states, the second sum is the analog of the first for particle 2, and the third term represents the connected diagrams. This interpretation would be exactly right if the momenta were quantized as $p_i = 2\pi n_i/L$ for integers $n_i$'s. This \emph{is} the case for the momentum $p_1$ of the magnon in the single-particle state, but not for $p_2$ and $p_3$.

The actual quantization of the momenta $p_2$ and $p_3$ in the two-particle state is determined by the Bethe equations \eqref{eqn:Bethe-equations}. Solving them iteratively for large $L$, we find
\<
  p_2 = \frac{2\pi n_2}{L} - \frac{4\pi}{L^2} \frac{n_2n_3}{n_2-n_3} + \order(L^{-3})
  \comma
  p_3 = \frac{2\pi n_3}{L} + \frac{4\pi}{L^2} \frac{n_2n_3}{n_2-n_3} + \order(L^{-3})
  \; .
\>
Now we can verify explicitly, that the sums in \eqref{eqn:ff-1to2-sums} produce Kronecker-deltas in the mode numbers at leading order in $1/L$, but that  they also give a subleading contribution when the mode numbers differ from each other, e.g.\
\<
  \sum_{x=1}^L e^{i(p_3-p_1)x} = \begin{cases}
    L + 2\pi i \frac{n_2n_3}{n_2-n_3}      & \text{for $n_1=n_3$} \; , \\[2mm]
      - \frac{2n_2n_3}{(n_1-n_3)(n_2-n_3)} & \text{for $n_1\not=n_3$} \; .
  \end{cases}
\>
The upshot is that also the sum-terms in \eqref{eqn:ff-1to2-sums} contain connected diagrams at a higher order in $1/L$. Now, we have three contributions
\<
  e^{ip_2} \sum_{1\leq x_3\leq L}e^{i (p_3-p_1) x_3} \eq \begin{cases}
    L + 2\pi i \, \frac{n_2^2}{n_2-n_3}          + \order(L^{-1}) & \text{for $n_1=n_3$} \; , \\[2mm]
      - \frac{2 n_2n_3}{(n_1-n_3)(n_2-n_3)}   + \order(L^{-1}) & \text{for $n_1\not=n_3$} \; ,
    \end{cases} \\[2mm]
  e^{ip_3} \Smat(p_3,p_2) \sum_{1\leq x_3\leq L}e^{i (p_2-p_1) x_3} \eq \begin{cases}
    L + 2\pi i \, \frac{(2n_2-n_3)n_3}{n_2-n_3}  + \order(L^{-1}) & \text{for $n_1=n_2$} \; , \\[2mm]
        \frac{2 n_2n_3}{(n_1-n_2)(n_2-n_3)}   + \order(L^{-1}) & \text{for $n_1\not=n_2$} \; ,
  \end{cases} \\[4mm]
  - e^{i(p_2+p_3-p_1)} \bigbrk{ 1 + \Smat(p_3,p_2) } \eq
      - 2   + \order(L^{-1}) \; .
\>
If all mode numbers are different, we obtain the connected piece of the form factor
\< \label{eqn:ff-1to2-conn}
 \bra{\psi(p_1)} S_{+,1} \ket{\psi(p_2,p_3)}_{\mathrm{conn}} \eq
  \frac{2 n_1(n_2+n_3-n_1)}{(n_1-n_2)(n_2-n_3)} \; .
\>
We will derive the same expression from the string theory result \eqref{eqn:ff-1to2-npp} in \secref{sec:matching-form-factors}. For a sensible comparison, we also have to divide by the norms of external states, however, to the order considered, these norms can be approximated by $||\psi(p_1)|| = \sqrt{L}$ and $||\psi(p_2,p_3)|| \approx L$ and will thus not contribute any momentum dependence.

\section{Relation between Spins and Strings}

In the study of the spectral problem, it was shown \cite{Kruczenski:2003gt} that the Landau-Lifshitz action describing the low-energy excitations about the ferromagnetic vacuum of the Heisenberg XXX spin-chain can be  matched to the string action in the fast string limit,  where one considers 
large charge strings, $J\rightarrow \infty$, and then expands to leading order in $\tilde \lambda =\lambda/J^2$.  This was extended in part to 
higher orders in $\tilde \lambda$ in subsequent works \cite{Kruczenski:2004kw, Kruczenski:2004cn}. As this simply reduces the string model to an alternative description of the spin-chain, it guarantees a matching of of all quantities, including those off-shell, at this order. Nonetheless it is useful to see how this explicitly works for the form factors and for the specific light-cone gauge choices on the worldsheet. 

\subsection{Mapping between Spin-Chain and Worldsheet Operators}

In this subsection, we will recall the dictionary between spin-chain and worldsheet fields. We again focus on the subsector described by the Heisenberg $\grSU(2)$ spin-chain and strings restricted to a $\Reals\times\Sphere^3$ subspace. The naive mapping would simply identify the spin raising operator, $S_{+} = S_1 + i S_2$, with the complex field, $Y = (Y_1+i Y_2)/\sqrt{2}$ (maybe up to a multiplicative constant). However, the actual mapping is non-linear and this naive map is only the leading term in a series expansion. One way to find the first subleading terms is to use the Landau Lifshitz model (non-relativistic sigma model on $\Sphere^2$), which is both the low energy effective field theory of the Heisenberg spin-chain and the sector of fast-moving strings on $\Reals\times\Sphere^3$.

In the Landau-Lifshitz description of the spin-chain, see e.g. \cite{Fradkin:1991nr, Affleck:1988nt}, the spin operators $\vec{S}_x$, acting on site $x=1,\ldots,L$, are replaced by their time-dependent expectation values in a coherent state, $\ket{n(\tau)}$, given by a unit 3-vector field $\vec{n}(\tau,\sigma)$ on a circle $\sigma\in\Sphere^1$ according to
\< \label{eqn:S-n-relation}
  \bra{n(\tau)} \vec{S}_x \ket{n(\tau)} = \Half \vec{n}(\tau, \sigma) \qquad \text{with $\sigma = 2\pi x/L$} \; .
\>
This becomes the Landau-Lifshitz field in the infinite volume, $L\rightarrow \infty$, limit and the link to the worldsheet field $Y$ is found by comparing the two parametrizations of the three-sphere that are used in the different contexts. In the string sigma model, the three-sphere is 
parametrized by $\vec{Y} = (Y_1,Y_2)^\trans$ and $\varphi$ as
\< \label{eqn:embedd-y1y2phi}
  X_1 + i X_2 = \frac{Y_1 - i Y_2}{1+\vec{Y}^2/4}
  \comma
  X_3 + i X_4 = \frac{1-\vec{Y}^2/4}{1+\vec{Y}^2/4} \, e^{i\varphi}
  \; ,
\>
where $X_1^2 + \ldots + X_4^2 = 1$ are embedding coordinates. To make contact with the Landau-Lifshitz model, a 
Hopf parametrization of the three-sphere is used \cite{Kruczenski:2003gt, Kruczenski:2004kw}
\< \label{eqn:embedd-u1u2alpha}
  X_1 + i X_2 = u_1 e^{i\alpha}
  \comma
  X_3 + i X_4 = u_2 e^{i\alpha}
  \; ,
\>
where $u_1$ and $u_2$ are complex and subject to the constraint $\abs{u_1}^2 + \abs{u_2}^2 = 1$. The angle $\alpha$ is real and introduces a gauge freedom which allows us to choose the phase of the vector $u = (u_1,u_2)^\trans$ arbitrarily. The vector $\vec{n}$ is related to these coordinates by
\< \label{eqn:n-from-u}
  \vec{n} = u^\dagger \vecsigma u \; .
\>
Note that the phase of $u$ drops out when we go to $\vec{n}$, so the relationship between $\vec{n}$ and $Y$ will not depend on this gauge freedom. From the above formulas, if follows that
\< \label{eqn:spin-string-map}
  S_+ \ \hat{=} \ \frac{n_1 + i n_2}{2} = u_1^* u_2 = \sqrt{2} \, \frac{1-\abs{Y}^2/2}{\bigbrk{1+\abs{Y}^2/2}^{2}} \, Y  e^{i\varphi} \;, 
\>
where we have denote the map from the operator to the expectation value by $\hat{=}$ and $S_-$ is the complex conjugate of this. 
The expansion in powers of the field reads
\< \label{eqn:spin-string-map-expanded}
  S_+ \ \hat{=} \ \sqrt{2} \, Y  e^{i\varphi} \lrsbrk{ 1 - \frac{3}{2} \abs{Y}^2 + \ldots } \; ,
\>
and displays the first correction to the naive dictionary which we are going to use in \secref{sec:matching-form-factors}.

\subsection{Mapping between Landau-Lifshitz and Near-Plane-Wave Model}

The relation between the variables, $S_\pm$, $n_\pm$, and $Y$, of the spin-chain, the Landau-Lifshitz model, and the near-plane-wave description of the $\grSU(2)$ sector goes beyond this kinematical relationship and also holds at the dynamical level as shown in \cite{Kruczenski:2003gt}. To see this correspondence for the light-cone gauge fixing used in the pertrubative calculation \secref{sec:perturbative-npw} we  need to know the appropriate gauge (value of the gauge parameter $a$) and, relatedly, the mapping between the spin-chain and the worldsheet lengths.

In terms of the field $\vec{n}(\tau,\sigma)$, the coherent state variable representing the spin-chain state in the thermodynamic limit, the Landau-Lifshitz action is given by
\cite{Fradkin:1991nr, Affleck:1988nt}
\<
  \Action = \int\!d\tau d\sigma\: \lrsbrk{  \Half \frac{ n_2 \dot{n}_1 - n_1 \dot{n}_2}{1+n_3} - \Quarter \acute{\vec{n}}^2 } \; ,
\>
where we preferred to write the Wess-Zumino term in a local form at the expense of breaking manifest $\grSO(3)$ invariance. The third component, $n_3$, is not an independent field, but rather given by $n_3 = \sqrt{1-n_1^2-n_2^2}$.

We can introduce the complex field $\phi = \half(n_1 + in_2)$, which corresponds to the spin-operator $S_+$. The action in terms of this field reads\footnote{It is possible and sometimes more convenient to bring the kinetic term into the standard form by working with a field $\hat{\varphi}$ that is related to $\phi$ by $\phi =\hat{ \varphi}\bigbrk{1-\abs{\hat{\varphi}}^2}^{1/2}$. Another nice feature is that the action in terms of $\hat{\varphi}$ will not have any interaction terms with time-derivatives.}
\< \label{eqn:LL-action-phi}
 \Action \eq \int\!d^2x\,\Biggsbrk{
     \frac{i(\phi^* \dot{\phi} - \dot{\phi}^* \phi)}{1+\bigbrk{1-4\abs{\phi}^2}^{1/2}}
   - \abs{\acute{\phi}}^2
   - \frac{(\phi^*\acute{\phi})^2+(\acute{\phi}^* \phi)^2}{1-4\abs{\phi}^2}
   - 2\,\frac{\abs{\phi}^2\abs{\acute{\phi}}^2}{1-4\abs{\phi}^2}
   } \; .
\>

Next, we will show that this action can also be obtained from the string action in the near-plane-wave limit by a number of redefinitions, which is essentially a simplified version of the relation given in \cite{Kruczenski:2003gt, Kruczenski:2004kw}. This matching will work in the $a=1$ gauge and only in this gauge.

Starting from the action in \eqref{eqn:agauge_pp_action}, we convert to the fields $y$ defined by $Y = y\, e^{-i \tau}$. Separating off the phase factor  allows
us to concentrate on fast moving strings. By computing
\<
\abs{\dot{Y}}^2 = \abs{\dot{y}}^2 + \abs{y}^2 + i ( y^* \dot{y} - \dot{y}^* y ) \; ,
\>
we see that this both removes the mass term and introduces first order time-derivatives. In order to take the fast spinning string limit, we rescaling the time coordinate $\tau$ by a parameter $\kappa$ and the space coordinate $\sigma$ by $\sqrt{\kappa}$, expand in $\kappa\rightarrow\infty$, and keep only the terms up to $\kappa^{-1}$. This amounts to discarding all terms with more then one $\tau$- and more than two $\sigma$-derivatives and leaves us with the non-relativistic action
\<
  \Lagr \eq \frac{1}{\kappa} \Biggsbrk{ i( y^* \dot{y} - \dot{y}^* y )
        - \abs{\acute{y}}^2
        + 2\abs{y}^2 \abs{\acute{y}}^2
        +\frac{1-2a}{2} \lrsbrk{
             (y^* \acute{y})^2 + (\acute{y}^* y)^ 2
           + 2i \abs{y}^2 \Bigbrk{ y^* \dot{y} - \dot{y}^* y }
        }
        } \nl + \order(\kappa^{-2}) \; .
\>
There are no terms of order $\kappa^0$. This is already quite similar to \eqref{eqn:LL-action-phi} but as is expected from \eqref{eqn:spin-string-map}, we need a non-linear field redefinition. In fact, by identifying $S_+$ with $\phi$, inserting $y\, e^{-i \tau}$ for $Y$, and fixing $\varphi$ to $\tau$, \eqref{eqn:spin-string-map} can tell us precisely the required redefinition, namely
\<
  y = \frac{1}{\sqrt{2}} \, \phi \lrbrk{ 1 + \frac{3}{4} \abs{\phi}^2 + \ldots } \; .
\>
We have truncated the series after the second term as this is sufficient to determine the action up to and including quartic interactions:
\<
\eval{\Lagr}_{\kappa^{-1}} \eq
           \iHalf ( \phi^* \dot{\phi} - \dot{\phi}^* \phi )
         - \Half \abs{\acute{\phi}}^2
         + \iHalf (2-a) \abs{\phi}^2 ( \phi^* \dot{\phi} - \dot{\phi}^* \phi ) \nl
         - \Quarter (1+a) \bigsbrk{ (\phi^* \acute{\phi})^2 + (\acute{\phi}^* \phi)^ 2 }
         - \abs{\phi}^2 \abs{\acute{\phi}}^2
         + \order(\phi^6) \; .
\>
Expanding out the Landau-Lifshitz action \eqref{eqn:LL-action-phi} to the same order, we observe agreement between the two models for
\<
  a = 1 \; ,
\>
where also a rescaling of the $\sigma$-coordinate such that $\partial_\sigma \to \sqrt{2} \, \partial_\sigma$ was necessary. This computation shows that when off-shell, gauge-dependent, quantities are supposed to be compared between the worldsheet and spin-chain descriptions, then it is most convenient to work in the $a=1$ gauge.

In fact, we can see that this gauge provides the most direct natural relation between the length of the worldsheet and the length of the spin-chain. According to \eqref{eqn:gauge-parameter}, in this gauge, the length of the worldsheet is given by to the string energy, $L=2\pi \mathcal{E}$. The string energy is in turn  given by the energy of the vacuum, given by the R-charge $J$, plus the sum of the fluctuation frequencies $\omega_i = \bigbrk{1+\lambda/J^2 n_i^2}^{1/2} = 1 + \order(J^{-2})$. For large $J$ the energy is thus equal to $J$ plus the number of excitations, $M$. And indeed, the spin-chain length is the sum of the up-spins $J$ plus the number of down-spins $M$.

\subsection{Matching Form Factors at Strong and Weak Coupling}
\label{sec:matching-form-factors}

While the matching of the actions ensures that the near-plane-wave form factor will match the spin-chain result in the appropriate limit
is also useful to see how this occurs explicitly.  To this end, we will show how \eqref{eqn:ff-1to2-conn} is reproduced from the string theory in the $a=1$ gauge using the map \eqref{eqn:spin-string-map-expanded}. Starting from the tree-level three-particle form factor in \eqref{eqn:tree_ff_0to3}, we first go to the crossed channel by sending $\vec{p}_1\to-\vec{p}_1$:
\<
\label{eqn:tree_ff_1to2}
  \bra{p_1} Y \ket{p_2,p_3} = - 2 \frac{ (p_2+p_3)^2 + (1-2a) (\vec{p}_1\cdot\vec{p}_{23\bar{1}} \,  \vec{p}_2\cdot\vec{p}_3 - 1) }{\sqrt{8\eps_1\eps_2\eps_3} \; (\vec{p}_{23\bar{1}}^2 - 1) } \; .
\>
Next, setting $p_i = 2\pi n_i/L$ and expanding for large $L$, we find at leading order 
\<
  \bra{p_1} Y \ket{p_2,p_3} = \frac{1}{2\sqrt{2}} \frac{ n_1(n_2+n_3-n_1) + 3 n_2 n_3 }{ (n_1-n_2)(n_1-n_3) } \; ,
\>
where we also set $a=1$. This expression already has the same poles as \eqref{eqn:ff-1to2-conn}, but the numerator still disagrees. However, if we add to this the string theory tree-level form factor
\<
  \bra{p_1} Y Y \bar{Y} \ket{p_2,p_3} = \frac{2}{\sqrt{8\eps_1\eps_2\eps_3}} = \frac{1}{\sqrt{2}} + \order(L^{-1}) \; ,
\>
with the coefficients predicted by \eqref{eqn:spin-string-map-expanded}, we do obtain
\< \label{eqn:ff-1to2-npp}
  \bra{p_1} \, \sqrt{2} \lrbrk{ Y - \frac{3}{2} Y Y \bar{Y} } \, \ket{p_2,p_3} = \frac{ 2 \, n_1(n_2+n_3-n_1) }{ (n_1-n_2)(n_1-n_3) } \; ,
\>
in agreement with the spin-chain result. This is quite analogous the matching found between the near-plane-wave 
and the one-loop spin-chain energies \cite{Parnachev:2002kk,Callan:2003xr,Callan:2004uv} and, as was found in that case, we 
expect it to fail at some sufficiently high order in the momentum expansion.

\section{Conclusions and Outlook}

We have formulated a set of consistency conditions for the form factors in the light-cone gauge fixed worldsheet theory for strings in $\AdS_5\times\Sphere^5$: a two-dimensional, massive, integrable, \emph{non-relativistic} field theory. These conditions are a straightforward generalization of Smirnov's axioms for relativistic theories, the main difference being that the worldsheet form factors depend on individual momenta whereas in the relativistic case, they are naturally functions of rapidity differences only. Focusing on an $\grSU(2)$ sector and working at tree-level in the near-plane-wave limit and up to one-loop in the near-flat-space limit, we computed the form factors for a single field, ${\cal O}=Y$, and for the simplest composite operator, ${\cal O}=\half\! :\!\!Y^2\!\!:$,  with various numbers of external particles in order to verify the proposed axioms. We also discussed the weak 't Hooft coupling or spin-chain limit of the worldsheet theory. Form factors for the Heisenberg spin-chain, as a consequence of similar properties of the Bethe states, quite naturally satisfy the same axioms although the spin-chain, being a lattice model, looks a priori quite different from the continuous worldsheet theory. 

Form factors are off-shell and gauge dependent quantities and therefore non-trivial to compare between string and spin-chain description. We demonstrated the non-linear map between worldsheet fields and spin-chain operators in an $\grSU(2)$ sector necessary to match the form factors on the two sides. We also showed that in the $a=1$ light-cone gauge, it is possible to compute the thermodynamic limit of the spin-chain form factors directly from the small momentum limit of the worldsheet theory. That being said, it is not expected that this match between one-loop gauge theory and string theory will persist to arbitrarily high order. But even if the quantitative relationship ceases to exist, this discussion 
emphasized how the form factors on the two sides and their computation are related on a conceptual level. Of course it would be very interesting to go beyond this limit and find an expression for the form factors which interpolated between weak and strong coupling. 

One possibility is to try to solve the proposed axioms directly. While the more general momentum dependence makes this task much 
harder than in the case of relativistic models, we still expect that the general 
strategy \cite{Weisz:1977ii, Karowski:1978vz, Smirnov:1992vz} may be applicable. An important example
should be the two-particle form factor $f^{\cal O}(z_1, z_2)$, which satisfies the functional equations
\<
f^{\cal O}(z_2, z_1)=f^{\cal O}(z_1, z_2)\Smat(z_1,z_2)=f^{\cal O}( z_1 +2\,\omega_2, z_2)
\>
where the $\Smat(z_1, z_2)$ is the exact S-matrix in the $\alg{su}(2)$ sector including the BES dressing phase \eqref{eq:exact_Smat}. We also expect that the structure of the form factors splits, similar to the relativistic case, in a part that is characteristic of the operator, a part that is characteristic of the external state, and a normalization. The perturbative calculations in this work should provide boundary conditions to the solution of these equations: helping to identify a solution with a given operator and fixing the normalization. 

It would of course be natural to  consider more general local operators, e.g.\ containing more fields, other flavors, derivatives or fermions. It is likely that a study of $\alg{psu}(2|2)^2\ltimes \mathbb{R}^3$ symmetries of the vacuum acting on the form factors will provide various relations between the varied form factors. Another possibility, though one that is challenging with our current understanding, is to try to generalize the off-shell Bethe ansatz 
methods \cite{Babujian:1990ii, Babujian:1993tm, Babujian:1993ts} to capture the vectorial aspects of the form factors. Of course perhaps the most interesting extension would be to  exponential operators which are likely appropriate for string vertex operators. This would be a way to try to make contact to the calculation of holographic correlation functions, in particular to the case involving two heavy and one light string, the semiclassical computation of which was recently initiated in \cite{Zarembo:2010rr,Costa:2010rz}.

In fact, it would be interesting to attempt a semiclassical analysis of the form factors of the worldsheet theory more generally, as this may give more insight into their exact expressions. Such semiclassical methods have been been previously developed for relativistic theories \cite{Goldstone:1974gf, Jackiw:1975im, Callan:1975yy, Mussardo:2003ji}, where for a scalar field theory with a potential, $V(\phi)$, such that is has kink solutions $\phi_{cl}(\sigma)$ the expectation value of a fundamental field between asymptotic kink states of mass $M$ and momenta $p_1$ and $p_2$ is given by,
\<
  \bra{p_1} \phi(0) \ket{p_2} = \int da~ e^{i  a (p_1-p_2)} \phi_{cl}(a)~.
\>
This equation is expected to hold to leading order in the coupling $\lambda$, where the kink mass is very large, scaling as $M\sim\tfrac{1}{\lambda}$, and the kink rapidities are very small $\theta_{1,2}\sim \lambda$. It may be worthwhile to find the generalization of this formula to the light-cone worldsheet theory as it does seem suggestive of the semiclassical heavy-heavy-light calculations.  

\section*{Acknowledgements}

We would like to thank the participants and organizers of the \textit{Common Trends in Gauge Fields, Strings and Integrable Systems} workshop at Nordita, and
in particular Z. Bajnok for his pellucid presentation on form factors in relativistic integrable models. We are also grateful to C. Ahn and S. Frolov for very useful suggestions. 

\appendix

\section{S-matrix in near-flat-space limit}
\label{sec:near-flat-S-mat}

In order to verify the form factor axioms in the perturbative computation, we need to know the one-loop two-particle S-matrix. It is most easily written down in terms of $\grSU(2|2)^2$ indices $A = (a|\alpha) = (1,2|3,4)$ and $\dot{A} = (\dot{a}|\dot{\alpha}) = (\dot{1},\dot{2}|\dot{3},\dot{4})$. There is a linear relation between the $\grSO(8)$ fields $(Y_{i'},Z_i)$ and $\psi$ and the $\grSU(2|2)^2$ fields $Y_{a\dot{a}}$, $Y_{\alpha\dot{\alpha}}$, $\Psi_{a\dot{\alpha}}$, and $\Bsi_{\alpha\dot{a}}$. The only explicit relation that we need here is
\<
 Y = \frac{1}{\sqrt{2}} \lrbrk{ Y_1 + i Y_2} = \frac{1}{\sqrt{2}} \, Y_{1\dot{1}}
 \comma
 \bar{Y} = \frac{1}{\sqrt{2}} \lrbrk{ Y_1 - i Y_2} = \frac{1}{\sqrt{2}} \, Y_{2\dot{2}}
 \; .
\>
The general two-particle worldsheet S-matrix has the index structure (compare e.g.\ \cite{Ogievetsky:1987vv}) 
\< \label{eqn:Smat-2p}
 \Smatrix_{A\dot{A}B\dot{B}}^{C\dot{C}D\dot{D}}(\eta_1,\eta_2) =
 (-)^{\abs{\dot{A}}\abs{B} + \abs{\dot{C}}\abs{D}} \,
 S_0(\eta_1,\eta_2) \,
 \smatrix_{AB}^{CD}(\eta_1,\eta_2) \,
 \smatrix_{\dot{A}\dot{B}}^{\dot{C}\dot{D}}(\eta_1,\eta_2) \; ,
\>
and the matrix part is usually parametrized as follows
\begin{align} \label{eqn:Smat-mat}
  \smatrix_{\lAA\lBB}^{\lCC\lDD} & = A \,\delta_\lAA^\lCC \delta_\lBB^\lDD
                                   + B \,\delta_\lAA^\lDD \delta_\lBB^\lCC \; , &
  \smatrix_{\lAA\lBB}^{\lcc\ldd} & = C \,\levi_{\lAA\lBB} \levi^{\lcc\ldd} \; , &
  \smatrix_{\lAA\lbb}^{\lCC\ldd} & = G \,\delta_\lAA^\lCC \delta_\lbb^\ldd \; , &
  \smatrix_{\lAA\lbb}^{\lcc\lDD} & = H \,\delta_\lAA^\lDD \delta_\lbb^\lcc \; , \\
  \smatrix_{\laa\lbb}^{\lcc\ldd} & = D \,\delta_\laa^\lcc \delta_\lbb^\ldd
                                   + E \,\delta_\laa^\ldd \delta_\lbb^\lcc \; , &
  \smatrix_{\laa\lbb}^{\lCC\lDD} & = F \,\levi_{\laa\lbb} \levi^{\lCC\lDD} \; , &
  \smatrix_{\laa\lBB}^{\lcc\lDD} & = L \,\delta_\laa^\lcc \delta_\lBB^\lDD \; , &
  \smatrix_{\laa\lBB}^{\lCC\ldd} & = K \,\delta_\laa^\ldd \delta_\lBB^\lCC \; . \nn
\end{align}
In the near-flat space limit, the prefactor $S_0$ can be written to order $\gamma^4$ as \cite{Klose:2007rz}
\< \label{eqn:Smat-pre}
  S_0(\eta_1,\eta_2) = \frac{\,\,e^{
  \frac{8i}{\pi} \gamma^2 \, \frac{\eta_1^3 \eta_2^3}{\eta_2^2 - \eta_1^2}
  \lrbrk{1-\frac{\eta_2^2 + \eta_1^2}{\eta_2^2 - \eta_1^2} \, \ln\frac{\eta_2}{\eta_1}}
  }}
  {1+\gamma^2 \, \eta_1^2 \eta_2^2 \lrbrk{\frac{\eta_2 + \eta_1}{\eta_2 - \eta_1}}^2} \; .
\>
and the \emph{exact} coefficient functions are given by\cite{Klose:2007rz}
\begin{align} \label{eqn:Smat-coeffs}
  A(\eta_1,\eta_2) & =  1 + i\gamma \, \eta_1\,\eta_2\,\frac{\eta_2-\eta_1}{\eta_2+\eta_1} \; , &
  B(\eta_1,\eta_2) & = -E(\eta_1,\eta_2) = 4i\gamma\, \frac{\eta_1^2\,\eta_2^2}{\eta_2^2-\eta_1^2} \; , \nn \\
  D(\eta_1,\eta_2) & =  1 - i\gamma \, \eta_1\,\eta_2\,\frac{\eta_2-\eta_1}{\eta_2+\eta_1} \; , &
  C(\eta_1,\eta_2) & =  F(\eta_1,\eta_2) = 2i\gamma\, \frac{\eta_1^{3/2}\,\eta_2^{3/2}}{\eta_2+\eta_1} \; , \\
  G(\eta_1,\eta_2) & =  1 + i\gamma \, \eta_1\,\eta_2 \; , &
  H(\eta_1,\eta_2) & =  K(\eta_1,\eta_2) = 2i\gamma\, \frac{\eta_1^{3/2}\,\eta_2^{3/2}}{\eta_2-\eta_1} \; , \nn \\[3mm]
  L(\eta_1,\eta_2) & =  1 - i\gamma \, \eta_1\,\eta_2 \; . \nn
\end{align}

\pdfbookmark[1]{\refname}{references}
\bibliographystyle{nb}
\bibliography{FormFactors}

\end{document}